\documentclass[aps,prx,showpacs,floatfix,twocolumn,superscriptaddress,longbibliography]{revtex4-2}
\usepackage{epsf}
\usepackage{epsfig}
\usepackage{amsmath}
\usepackage{amsfonts}
\usepackage{amssymb}
\usepackage{graphicx}
\usepackage{multirow}
\usepackage{graphicx}
\usepackage[table,xcdraw]{xcolor}
\usepackage{color}
\usepackage{bm}
\usepackage[yyyymmdd]{datetime}
\usepackage{dashrule}
\usepackage{ulem}
\usepackage{pifont}
\usepackage{lineno}
\usepackage[margin=0.7in]{geometry}

\newcommand{\vb}[1]{\textcolor{black}{#1}}
\newcommand{\jl}[1]{\textcolor{black}{#1}}
\newcommand{\bi}[1]{\textit{\textbf{#1}}}

\newcommand{\com}[1]{\textcolor{black}{#1}}

\begin{document}

    \title{\com{Anisotropic electron damping and energy gap in Bi$_2$Sr$_2$CaCu$_2$O$_{8+\delta}$}}

    \author{Jiemin Li}
    \email{jli1@bnl.gov}
    \affiliation{National Synchrotron Light Source II, Brookhaven National Laboratory, Upton, NY 11973, USA}
    \author{Yanhong Gu}
    \affiliation{National Synchrotron Light Source II, Brookhaven National Laboratory, Upton, NY 11973, USA}
    \author{Takemi Yamada}
    \affiliation{Liberal Arts and Sciences, Toyama Prefectural University, Imizu, Toyama 939-0398, Japan}
    \author{Zebin Wu}
    \affiliation{Condensed Matter Physics and Materials Science Department, Brookhaven National Laboratory, Upton, NY, 11973, USA}
    \author{Genda Gu}
    \affiliation{Condensed Matter Physics and Materials Science Department, Brookhaven National Laboratory, Upton, NY, 11973, USA}
    \author{Tonica Valla}
    \affiliation{Condensed Matter Physics and Materials Science Department, Brookhaven National Laboratory, Upton, NY, 11973, USA}
    \affiliation{\com{Donostia International Physics Center, E-20018 Donostia-San Sebastian, Spain}}
    \author{Ilya Drozdov}
    \affiliation{Condensed Matter Physics and Materials Science Department, Brookhaven National Laboratory, Upton, NY, 11973, USA}
    \author{Ivan Bo\v{z}ovi\'{c}}
    \affiliation{Condensed Matter Physics and Materials Science Department, Brookhaven National Laboratory, Upton, NY, 11973, USA}
    \author{Mark P. M. Dean}
    \affiliation{Condensed Matter Physics and Materials Science Department, Brookhaven National Laboratory, Upton, NY, 11973, USA}
    \author{Takami Tohyama}
    \email{tohyama@rs.tus.ac.jp}
    \affiliation{Department of Applied Physics, Tokyo University of Science, Katsushika, Tokyo 125-8585, Japan}
    \author{Jonathan Pelliciari}
    \email{pelliciari@bnl.gov}
    \affiliation{National Synchrotron Light Source II, Brookhaven National Laboratory, Upton, NY 11973, USA}
    \author{Valentina Bisogni}
    \email{bisogni@bnl.gov}
    \affiliation{National Synchrotron Light Source II, Brookhaven National Laboratory, Upton, NY 11973, USA}
    
    \begin{abstract}
        The many body electron-electron interaction in cuprates causes the broadening of the electronic bands in \textit{\textbf{k}}-space, leading to a deviation from the standard Fermi liquid. While a \textit{\textbf{k}}-dependent anisotropic electronic scattering (\textit{\textbf{k}}-DAES) has been assessed by photoemission, its fingerprint in \textit{\textbf{Q}}-space has been scarcely considered. Here, we explore the \textit{\textbf{Q}}-dependent electron dynamics in optimally doped Bi$_2$Sr$_2$CaCu$_2$O$_{8+\delta}$ through the evolution of low-energy charge excitations as measured by resonant inelastic x–ray scattering (RIXS). In the normal state, the RIXS spectra display a continuum of excitations down to 0~meV, while the superconducting state features a spectral weight suppression below 80 meV without any enhancement at higher energies. To interpret the  energy and \textit{\textbf{Q}}-evolution of our data, we introduce a phenomenological expression of the charge susceptibility by including the \textit{\textbf{k}}-DAES. We show that only the charge susceptibility with \textit{\textbf{k}}-DAES captures the RIXS data, highlighting the importance of \textit{\textbf{k}}-DAES when describing the \textit{\textbf{Q}}-dependence of charge excitations from 0 to few eV scale. Furthermore, we also find that the inclusion of \textit{\textbf{k}}-DAES is essential when quantitative parameters such as the \com{electronic energy} gap are extracted from RIXS data.
    \end{abstract}

    \maketitle

    \textit{Introduction.}---In Bardeen–Cooper–Schrieffer theory, the electron-phonon coupling generates an effective attractive interaction between two electrons to form the so-called Cooper pairs, whose condensation at low-temperature gives rise to superconductivity~\cite{Bardeen1957}. This concept however fails to account for unconventional superconductivity in cuprates~\cite{Bednorz1986} that appears at a higher temperature than what the electron-phonon coupling can support~\cite{Nagamatsu2001}. Instead, strong electron-electron (\textit{el-el}) interaction features in these unusual cases and is believed to play a predominant role for the formation of Cooper pairs~\cite{Keimer2015}. In addition, it has also been argued that the \textit{el-el} interaction is at the root of multiple exotic phases of cuprates, such as magnetism, charge/spin density waves, pseudogap, and strange metal behaviors~\cite{Keimer2015}. Therefore, the understanding of \textit{el-el} interaction is crucial to achieve a physical understanding of these phenomena and of unconventional superconductivity.

    Generally, such strong \textit{el-el} interaction can significantly alter the electron dynamics in cuprates, leading to the deviation from a Fermi liquid. Indeed, Raman scattering or infrared reflectivity experiments revealed a strong electron damping, i.e. an anomalous scattering rate that displays a linear variation with temperature and electron energy~\cite{Thomas1988,Collins1989,Ivan1991,Cooper1992,Elihu2000} and that is suppressed in the superconducting (SC) state~\cite{Bonn1992, Rieck1995}. These unusual behaviors could be partially described by the theory of ``marginal Fermi liquids"~\cite{Varma1989}, which is however uncapable of explaining the anisotropy of the scattering rate~\cite{Abdel2006, Fang2022} in \textit{\textbf{k}} space, as reported by the angle-resolved photo-emission spectroscopy (ARPES)~\cite{Valla1999,Valla2000,Kaminski2005,Chang2013}. While the origin of the \textit{\textbf{k}}-dependent anisotropic electron scattering (\textit{\textbf{k}}-DAES) is still the subject of debate, two contributions to the anomalous electron scattering in cuprates seem clear: (1) anisotropic elastic scattering possibly associated with impurity scattering~\cite{Varma2001, Elihu2000} or the pseudogap~\cite{Kaminski2005}; (2) inelastic part arising from the spin-fluctuation scattering~\cite{Norman1990,Millis1992,Dahm2005}.

        \begin{figure*}
            \includegraphics[width=0.95\textwidth]{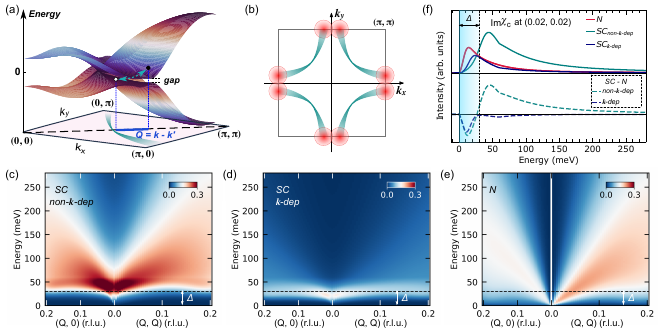}
            \caption{Sketch of particle-hole excitations across the energy gap in \textit{\textbf{k}}-space and the calculation of charge susceptibility $\chi_c(\textit{\textbf{Q}}, \omega)$. (a) A particle-hole excitation at \textit{\textbf{Q}} = \textit{\textbf{k}} $-$ \textit{\textbf{k}}$^{'}$, stemming from electron hopping from the lower band to the upper one along the nodal-$(Q, Q)$ direction. The bottom plane outlines a quadrant of the Brillioun zone of cuprates. The shaded arc on the bottom plane represents a simplified Fermi surface of cuprates, while the width denotes the anisotropy of $\Gamma^\mathrm{SC}_{\bi{Q},\bi{k}}(\omega)$ in the antinodal (Cu-O) region. The black dashed line indicates the nodal (diagonal Cu-Cu) direction. (b) Representation of the anisotropic \jl{\textit{\textbf{k}}-dependence} of $\Gamma^\mathrm{SC}_{\bi{Q},\bi{k}}(\omega)$ considered in the calculation of  Im$\chi_c(\textit{\textbf{Q}}, \omega)$ in the SC state. Eight Gaussian curves (indicated by red circles) centered at anti-nodal points are used to model the anisotropic momentum dependence. (c, d) Im$\chi_c(\textit{\textbf{Q}}, \omega)$ along nodal and anti-nodal directions in the SC state, without (c) and with (d) the anisotropic \jl{\textit{\textbf{k}}-dependence} in $\Gamma^\mathrm{SC}_{\bi{Q},\bi{k}}(\omega)$. Im$\chi_c(\textit{\textbf{Q}},\omega)$ in (d) was obtained for $\Gamma^\mathrm{SC}_{\bi{Q},\bi{k}}(\omega)$ derived from $A$=6 and $\sigma$=0.45$\pi$, see Eq.~1. (e) Im$\chi_c(\textit{\textbf{Q}}, \omega)$ in the normal state. (f) Im$\chi_c(\textit{\textbf{Q}}, \omega)$ at (0.02, 0.02). The teal and blue dashed lines are the spectral difference between the normal and SC states without and with the \textit{\textbf{k}}-DAES respectively. The black dashed lines in (c, d, e, f) indicate the \com{electronic energy} gap size ($\sim$~30~meV) of nearly optimally dopped Bi2212.}
            \label{Fig1}
        \end{figure*}

    Nonetheless, the impact of \textit{\textbf{k}}-DAES on the cuprate electron dynamics has been scarcely discussed in \textit{\textbf{Q}} space~\cite{Mitrano2018, Husain2019}. It is however important to experimentally address this point to refine models of the electron response at finite \textit{\textbf{Q}}. Here, we examine the electron dynamics of optimally doped Bi$_2$Sr$_2$CaCu$_2$O$_{8+\delta}$ (Bi2212) using resonant inelastic x-ray scattering (RIXS), a momentum resolved two-particle probe~\cite{Ament2011, Mitrano2024exploring}. We study the low-energy charge excitations across the \com{electronic energy} gap of Bi2212 using high-resolution RIXS at the Cu $L_3$-edge, below and above $T_c$. In the normal (N) state, we observe a continuum of excitations down to $\sim$~0~meV. In the SC state, a spectral suppression 
    emerges up to 80~meV and displays a momentum dependence, consistent with the opening of the \com{electronic energy} gap. However, no spectral weight enhancement manifests at higher energies ($\gtrsim80$ meV), contradicting expectations based on spectral weight conservation. To explain our observations, we model the RIXS charge response in the SC state with the charge susceptibility formulated as a function of \textit{\textbf{k}}-DAES. We found that only  charge susceptibility including \textit{\textbf{k}}-DAES captures the data and their momentum dependence. Our results prove that besides capturing the opening of the \com{electronic energy} gap~\cite{Marra2013,Suzuki2018,Merzoni2024}, we can extract the electron dynamics including the damping in reciprocal space. This aspect therefore needs to be accounted for when quantifying the magnitude of the \com{electronic energy} gap.

    To start, we need to formulate a calculation of the charge susceptibility $\chi_c(\textit{\textbf{Q}}, \omega)$. Distinct from single-particle probes, such as ARPES, RIXS is a two-photon process, requiring the consideration of both occupied and unoccupied bands to describe the created particle-hole excitations.
    Figure~\ref{Fig1}(a) depicts a specific particle-hole excitation across the \com{energy} gap with $d$-wave symmetry along the nodal direction and exchanged momentum \textit{\textbf{Q}} = \textit{\textbf{k}} $-$ \textit{\textbf{k}}$^{'}$. The RIXS response for charge excitations is the combination of all particle-hole pairs satisfying the momentum conservation \textit{\textbf{Q}} = \textit{\textbf{k}} $-$ \textit{\textbf{k}}$^{'}$. \textcolor{black}{We approximate the RIXS response of charge excitations as proportional to the charge dynamic structure factor, $I^\mathrm{RIXS}\propto S_c(\textit{\textbf{Q}}, \omega)$ = Im$\chi_c(\textit{\textbf{Q}}, \omega)/[1-e^{-\omega/(k_BT)}]$~\cite{Ament2011, Marra2013, Jia2016}, where \textit{\textbf{Q}} and $\omega$ are respectively the momentum and energy transfer from the photon to the material. The imaginary of charge susceptibility Im$\chi_c(\textit{\textbf{Q}}, \omega)$} is extracted from a tight binding model as reported in Eqs.~(S1) and (S2) of Supplementary Material \cite{sm} for SC or N state. Below we focus on \jl{the momentum \bi{k}-dependent scattering rate $\Gamma^\mathrm{SC}_{\bi{Q},\bi{k}}(\omega)$ for a given \bi{Q}} as it is the dominant physical quantity in $\chi_c(\textit{\textbf{Q}}, \omega)$ that is sensitive to the electron-electron interaction in the SC state:
    \jl{
    \begin{equation}
        \begin{aligned}[t]
            \Gamma^\mathrm{SC}_{\bi{Q},\bi{k}}(\omega) 
           = \Biggr[\frac{\omega^n}{(2\Delta)^{n-1}} H(2\Delta-\omega)+\omega H(\omega - 2\Delta)
              \Biggr] \\
               \times \Biggr[1 + A\sum_{i=1}^{4}\frac{e^{-|\textit{\textbf{k}}-\textit{\textbf{k}}_i|^2/(2\sigma^2)} + e^{-|\bi{k}+\bi{Q}-\textit{\textbf{k}}_i|^2/(2\sigma^2)}}
               {\sqrt{2\pi}\sigma} \Biggr].
        \end{aligned}
        \label{Eq2}
    \end{equation}  
    }
    In the above equation, the first square bracket contains the Heaviside step function $H[x]$  and refers to the energy dependence of the scattering rate separated by the \com{energy gap $2\Delta$ with a $d$-wave symmetry that primarily originates from the superconducting phase but also partially from the pseudo-gap~\cite{Norman1990, Norman1995, Lee2007, Kondo2009}}. The scattering rate is assumed to follow an $\omega$-linear marginal Fermi liquid form outside the gap and an $\omega^n$ power law (with $n=2$) inside the gap expressing a suppression of the rate due to the gap. Note that the choice of a larger value of \textit{n}, for example $n=3$,  does not change our conclusions. The second term phenomenologically describes the anisotropic \textit{\textbf{k}}-dependence of $\Gamma^\mathrm{SC}_{\bi{Q},\bi{k}}(\omega)$~\cite{Valla1999,Valla2000,Kaminski2005,Abdel2006,Chang2013,Fang2022} \textcolor{black}{which captures the pseudogap physics in cuprates and are} overall represented by \jl{eight Gaussians at the four anti-nodal points $\textit{\textbf{\textit{\textbf{k}}}}_i$}, see the red circles in the Bi2212 Fermi surface illustrated in Fig.~\ref{Fig1}(b). The pre-factor $A$ and the width $\sigma$ respectively characterize the magnitude and the extent of the anisotropy in momentum space. \com{This form of $\Gamma^\mathrm{SC}_{\bi{Q},\bi{k}}(\omega)$ successfully captures the essential behavior of the electron dynamics in cuprates as reported inprevious studies~\cite{Valla2000, Valla2020, Vishik2009, Vishik2012}.} With such formulation, we can thus switch on ($A \neq 0$) or off ($A=0$) the anisotropic \textit{\textbf{k}}-dependence of scattering rate for the calculation of Im$\chi_c(\textit{\textbf{Q}}, \omega)$ in the SC state, evaluating its impact on the RIXS cross-section.
    
        \begin{figure}[t]
            \includegraphics[width=0.7\columnwidth]{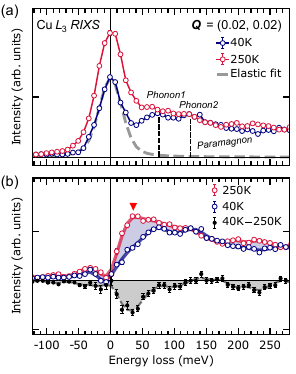}
            \caption{Cu $L_3$ RIXS spectra at (0.02, 0.02). (a) Data collected below (40 K, blue dotted line) and above (250 K, red dotted line) $T_C$. The dashed grey curve represents the elastic peak at 40 K modeled with a Voigt line shape. (b) RIXS spectra after elastic subtraction. The black dots refer to the spectral difference between the SC and normal states. The lines here correspond to the smoothed data. The error bars in the RIXS spectra are determined assuming a Poisson statistics, while the error bars in spectral difference are the sum of those from each RIXS spectrum.}
            \label{Fig2}
        \end{figure}

        \begin{figure*}
            \includegraphics[width=0.85\textwidth]{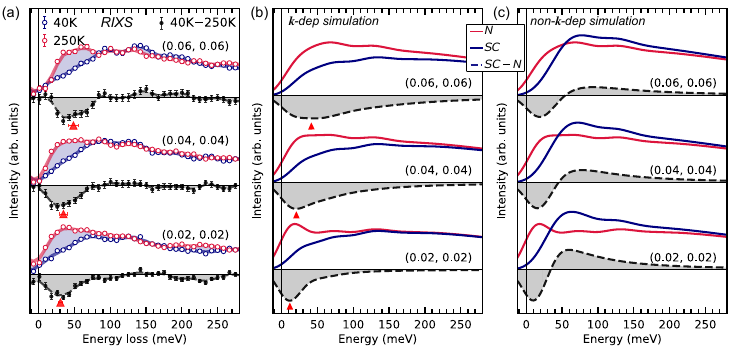}
            \caption{Momentum dependence of RIXS spectra in the normal and SC states, and the corresponding simulations along the nodal direction. (a) RIXS spectra after the subtraction of elastic and difference between the SC and N states spectra. The (dashed) lines are smoothed results. The error bars are defined as described in Fig.~\ref{Fig2}. (b) and (c) are the simulations calculated with and without the \textit{\textbf{k}}-DAES. \textcolor{black}{The red triangles in (a) and (b) respectively characterize the center of mass or the position of the dip features in the spectral differences. \com{We note that all calculated $SC$-$N$ curves present a weak spectral weight up to $\sim200$~meV, which is not captured in the experimental data. We associate this with the low sensitivity of high-resolution RIXS to broad and weak signals, \textcolor{black}{or possibly with other weak overlapping excitations.}}}
            }
            \label{Fig3}
        \end{figure*}

    Figures~\ref{Fig1}(c) and (d) show the calculated Im$\chi_c(\textit{\textbf{Q}}, \omega)$ in momentum-energy space for the SC state \jl{with \com{$\Delta=30$}~meV}, without and with the anisotropic \textit{\textbf{k}}-dependence of $\Gamma^\mathrm{SC}_{\bi{Q},\bi{k}}(\omega)$. Hereinafter, we refer to these cases as {\it non-\textit{\textbf{k}}-dependent} and {\it \textit{\textbf{k}}-dependent}, respectively. We focus on two reciprocal space directions: the anti-nodal $(Q, 0)$ and nodal $(Q, Q)$. In the non-\textit{\textbf{k}}-dependent case, see Fig.~\ref{Fig1}(c), the opening of the \com{energy} gap pushes the spectral weight of the charge excitations above the \com{$\sim \Delta$} for any \textit{\textbf{Q}} directions, leading to  a sharp intensity enhancement peaked at \com{$\sim 1.5\cdot\Delta$} around the Brillouin Zone (BZ) center~\cite{Suzuki2018}. In this work, we call this enhancement as a hump structure~\cite{Onari2010}. When switching on the anisotropic \textit{\textbf{k}}-dependence of $\Gamma^\mathrm{SC}_{\bi{Q},\bi{k}}(\omega)$, see Fig.~\ref{Fig1}(d) ($A$=6, $\sigma$=0.45$\pi$), the hump structure gets heavily suppressed, and overall the spectral weight along both anti-nodal $(Q, 0)$ and nodal $(Q, Q)$ directions weakens and shifts towards lower energies than the non-\textit{\textbf{k}}-dependent case.
    We note that, however, for large \textit{\textbf{Q}} values ($|$\textit{\textbf{Q}}$|\gtrsim$ 0.15) the spectral shapes of Im$\chi_c(\textit{\textbf{Q}}, \omega)$ for both cases become broad and indistinguishable besides an overall scaling factor. For completeness, we report in Fig.~\ref{Fig1}(e) the Im$\chi_c(\textit{\textbf{Q}}, \omega)$ in the normal state, where the scattering rate $\Gamma^\mathrm{N}(\omega)$ is approximated as a linear function of the electron energy $\omega$ following the marginal Fermi liquid form~\cite{Varma1989} (See Sec. II of the Supplementary Material~\cite{sm}). Strong charge excitations appear in both directions, ungapped at the BZ center and linearly dispersing in energy with an increasing width in the high-\textit{\textbf{Q}} regions. 

    To better visualize how Im$\chi_c(\textit{\textbf{Q}}, \omega)$ varies between the SC and N states, we take the calculations at \textit{\textbf{Q}}=(0.02, 0.02) [Figs.~\ref{Fig1}(c-e)] and compare them with  spectral differences {\it{SC-N}} (dashed lines) [Fig.~\ref{Fig1}(f) (solid lines)] . 
    In the non-\textit{\textbf{k}}-dependent scenario, the calculated {\it{SC-N}} difference yields a characteristic {\it{dip-peak}}-like feature (green dashed line), owing to the formation of the hump structure in the SC state at energies larger than \com{$\Delta$} ~\cite{Suzuki2018}. In the \textit{\textbf{k}}-dependent scenario, instead, the {\it{SC-N}} yields a simpler {\it{dip}}-like feature with a minimum below \com{$\Delta$}.  In Fig.~S1~\cite{sm} we display the complete {\it{SC-N}} map in reciprocal space. We  note a stronger contrast of {\it{SC-N}} along the nodal direction up to \textit{\textbf{Q}}$\sim$(0.1,0.1) rather than along the anti-nodal direction. Therefore, by experimentally investigating the {\it{SC-N}} spectral difference along the nodal direction and as a function of \textit{\textbf{Q}} \com{around the BZ center}, we aim at unravelling the electron dynamics in Bi2212 with RIXS. To this purpose, high-resolution RIXS is required to access the spectrum in the energy scale of \com{$\Delta$}.

    A high-quality single crystal of nearly optimally doped Bi$_2$Sr$_2$CaCu$_2$O$_{8+\delta}$ ($T_c$=91K, \com{$T^*$$\sim$190K})~\cite{Hashimoto2014} was used for this study. The RIXS experiment was performed at the SIX 2-ID beamline of the National Synchrotron Light Source II ~\cite{Dvorak16}, with an energy resolution of $\Delta E$ $\sim$ 35~meV (full width at half maximum) at the Cu $L_3$ edge. RIXS spectra were measured at three \textit{\textbf{Q}} points with an incident photon energy tuned to the maximum of the Bi2212 Cu $L_3$ absorption peak. \com{A previous RIXS study~\cite{Suzuki2018} demonstrated that the temperature dependence of the RIXS cross-section is sensitive to both $T_c$ and $T^*$, therefore to maximize the spectral contrast in our experimental data, we focused on two extreme conditions, $T < T_c$ (40 K) and $T>T^*$ (250 K).} \com{All spectra presented here are normalized to the integrated spectral weight in the region 1.0~$\sim$~4.0~eV \cite{Merzoni2024}}. \textcolor{black}{The zero-energy of the RIXS spectra was determined by fitting the spectrum within the region -120~$\sim$~20~meV with a Voigt function whose width was constrained to the energy resolution.} Further details on the sample and experimental configuration can be found in Sec. I of the Supplementary Material \cite{sm}. 
    
    Figure~\ref{Fig2}(a) reports 
    the RIXS spectra at \textit{\textbf{Q}}=(0.02, 0.02) in the SC (blue open dots) and normal (red open dots) states. In the low energy region, we can identify multiple excitations, previously discussed in RIXS studies of Bi-based cuprates~\cite{Dean2013, Dean2014, Dean2015, Chaix2017, Jiemin2020}. The peaks at $\sim$ 80 meV and $\sim$ 125 meV correspond, respectively, to the apical phonon (labelled as {\it{Phonon1}}) and a combined phonon mode (labelled as {\it{Phonon2}}) between the apical and the A$_{1g}$ phonons ~\cite{Devereaux2016}. At higher energies, the broad spectral weight stretching even beyond 250 meV is associated with the paramagnon mode~\cite{Dean2013, Chaix2017, Jiemin2020}. While these excitations do not show relevant temperature dependence between 40~K and 250~K, we observe instead a suppression of the spectral weight in the SC state, affecting the elastic energy range up to almost the first phonon mode. The elastic peak variation between 40~K and 250~K and its momentum evolution are discussed in Sec.~V of the Supplementary Material\jl{~\cite{sm}}, showing a strong connection with \com{the opening of the energy gap in SC state}. Here, we focus on the quasi-elastic portion of the RIXS data, see Fig.~\ref{Fig2}(b) after elastic peak removal. The spectral weight below $\sim$80~meV is clearly suppressed at 40~K with respect to 250~K. Such a behaviour is incompatible with a pure thermal effect, as it would broaden the high temperature data. Our observation thus suggests an electronic origin connected with the crossing of $T_c$ behind this response, which resembles the electron redistribution caused by the gap opening, see Fig.~\ref{Fig1}(f) and Ref.~\cite{Suzuki2018}. 
    
    To assess the spectral variation between SC and N states versus \textit{\textbf{Q}}, we introduce the difference spectrum 40~K-250~K, see the black dotted line in Fig.~\ref{Fig2}(b). Figure~\ref{Fig3}(a) presents data at \textit{\textbf{Q}}= (0.02, 0.02), (0.04, 0.04) and (0.06, 0.06) using the same format introduced in Fig.~\ref{Fig2}(b).
    A clear dip-like shape is observed in the 40~K-250~K spectra at the various \textit{\textbf{Q}} points. The dip is centered around $\sim30$~meV at \textit{\textbf{Q}}= (0.02, 0.02), and its center of mass quickly shifts towards higher energies, i.e. $\sim50$~meV, as \textit{\textbf{Q}} is increased to (0.06, 0.06) (see red triangles in Fig.~\ref{Fig3}(a)).
    
     \begin{figure}[t]
            \includegraphics[width=0.7\columnwidth]{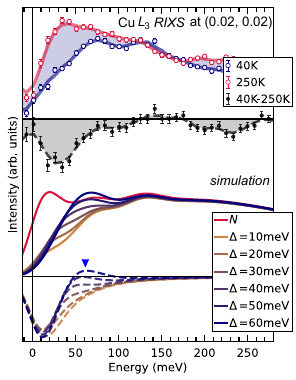}
            \caption{RIXS spectra at \textit{\textbf{Q}}=(0.02,0.02) and simulations derived from different \com{$\Delta$} values. The \textit{\textbf{k}}-DAES was fixed for all simulations and generated from $A$=6 and $\sigma$=0.45$\pi$. \textcolor{black}{The blue triangle indicates the hump feature emerging in the calculated spectral difference for \com{$\Delta>$}~40~meV.}
            }
            \label{Fig4}
        \end{figure}

    To comprehend this observation, we compare the RIXS data to the simulated spectra obtained from a model composed of two Gaussian peaks accounting for {\it{Phonon1}} and {\it{Phonon2}}, a heavily damped harmonic oscillator~\cite{Robarts2019} accounting for the paramagnon, and \textcolor{black}{the charge dynamic structure factor $S_c(\textit{\textbf{Q}}, \omega)$} accounting for the low-energy charge response, all convoluted with the instrumental resolution. Details about the model are presented in Sec.~III of the Supplementary Material \cite{sm}. For the Im$\chi_c(\textit{\textbf{Q}}, \omega)$ calculations, we further consider the non-\textit{\textbf{k}}-dependent and \textit{\textbf{k}}-dependent scattering rates for the SC state, and only one case for the N state.

    When the \textit{\textbf{k}}-dependent scattering rate is included, see Fig.~\ref{Fig3}(b), the calculated $SC$-$N$ spectra (black dashed line) display a dip-like shape with a momentum dependent dip position that moves towards higher energies, i.e., $\sim$~60~meV when \textit{\textbf{Q}} is increased to (0.06, 0.06). In the non-\textit{\textbf{k}}-dependent case, see Fig.~\ref{Fig3}(c), the calculated $SC$-$N$ spectra display a dip-peak-like shape with a \textit{\textbf{Q}}-dependent nodal point moving to higher energies for larger \textit{\textbf{Q}}s. 
    From these simulated spectra, we can exclude a phononic or magnetic origin for the \textit{\textbf{Q}}-dependent behaviour of $SC$-$N$, given that phonons and paramagnons have negligible energy dispersion within the investigated \textit{\textbf{Q}}-range ~\cite{Jiemin2020, Dean2013}. Rather, it indicates a sizeable contribution from the charge response to the overall low-energy spectral weight. In fact, the simulated $SC$-$N$ difference spectra in Figs.~\ref{Fig3}~(c,~d) and their momentum dependence come primarily from intrinsic changes in Im$\chi_c(\textit{\textbf{Q}}, \omega)$ rather than changes in the Bose factor, see Figs.~\ref{Fig1}(c-f) and Fig.~S1 in~\cite{sm}.
    
    By comparing the data in Fig.~\ref{Fig3}(a) with the simulations, it emerges that the \textit{\textbf{k}}-dependent scenario captures our main observations: (1) there is no trace of the hump structure; (2) the difference spectra match the simulations in Fig.~\ref{Fig3}~(b) in terms of spectral shape and momentum-dependence. The parameter optimization used in Fig.~\ref{Fig3}(b) is presented in Sec.~VI of the Supplementary Material~\cite{sm} where the pair of ($A$, $\sigma$) was selected to simultaneously satisfy our RIXS data as well as ARPES data \cite{Valla1999,Valla2000}. From a quantitative point of view, we also obtain a good agreement between the measured and calculated $SC$-$N$ dip-position at \textit{\textbf{Q}}=(0.04, 0.04), and (0.06, 0.06). However, we overestimate it at \textit{\textbf{Q}}=(0.02, 0.02), likely due to its proximity to the subtracted quasi-elastic peak. Indeed, as discussed in Sec.~V of the Supplementary Material -- where we examine the raw spectra before any elastic subtraction -- we observe a stronger intensity variation of the 40~K-250~K spectrum in correspondence to the quasi elastic line at \textit{\textbf{Q}}=(0.02, 0.02) as compared to slightly larger \textit{\textbf{Q}}s, see Fig.~S5. The same observation was found on a different cuprate material, La$_{1.85}$Sr$_{0.15}$CuO$_4$, see Fig.~S6. We interpret these findings as the direct fingerprint of the gap closure at \textit{\textbf{Q}}$\sim$0, thus the spectral distribution in the normal state concentrates below $\Delta$, partially mixing with the elastic line of the RIXS spectra.
    
    Our analysis demonstrates the need of including the \textit{\textbf{k}}-DAES to describe the \textit{\textbf{Q}}-space electron dynamics probed by RIXS. This enables us to examine the sensitivity of RIXS to the magnitude of $\Delta$. Figure~\ref{Fig4} displays the RIXS spectra and the simulations for different values of \com{$\Delta$}  at \textit{\textbf{Q}}= (0.02, 0.02). As discussed in Fig.~\ref{Fig1}, the spectral weight is pushed at higher energy than \com{$\sim \Delta$} in presence of the \textit{\textbf{k}}-DAES in the SC state, thus causing the $SC$-$N$ line to assume a spectral shape depending on \com{$\Delta$}. Comparing the calculated $SC$-$N$ spectra to our data, a good agreement can be reached when \com{$\Delta \lesssim$~40~meV}, because a dip-peak like shape comes out for larger \com{$\Delta$} values \textcolor{black}{(see the hump feature indicated by a blue triangle in Fig.~\ref{Fig4})}, contrary to the experimental observations that present only a dip. This result is consistent with the \com{$\Delta$} size of optimally doped Bi2212 ($\sim$~30~meV), as known from other studies~\cite{Vishik2012}. \textcolor{black}{Note that the spectral difference $SC$-$N$ for smaller \com{$\Delta$} comes from the \textit{\textbf{k}}-DAES, more than the presence of the \com{energy} gap.} Thus, it is necessary to account for the proper electron behaviour, i.e. the \textit{\textbf{k}}-DAES in the case of cuprates, when extracting the magnitude of \com{$\Delta$} from RIXS data.
    
    \vb{At last, we comment that the impact of \textit{\textbf{k}}-DAES stretches well above the superconducting gap energy scale, affecting charge excitations up to few eV as covered by Im$\chi_c(\textit{\textbf{Q}}, \omega)$ at the BZ boundary. This suggests that \textit{\textbf{k}}-DAES in the SC state should be included in general when modelling charge excitations in cuprates, e.g. temperature dependence of plasmonic excitations.}
        
    In summary, we investigated the low-energy charge excitations of optimally doped Bi2212 across the SC gap with RIXS. By comparing spectra measured in the SC and normal states at different \textit{\textbf{Q}} points close to the BZ center, we find a suppression of spectral weight in the SC state below $\sim80$ meV, without observing any enhancement at higher energies. The extracted $SC$-$N$ difference spectra, shaped as a dip centered around \com{$\Delta$} at \textit{\textbf{Q}}= (0.02, 0.02), further display a shift towards higher energies as \textit{\textbf{Q}} is increased. Such observations are well captured by charge susceptibility calculations when a \textit{\textbf{k}}-DAES is included. These results demonstrate the sensitivity of RIXS to the intrinsic electron dynamics of the cuprate superconductors, and more broadly the impact of the \textit{\textbf{k}}-DAES in \textit{\textbf{Q}}-sensitive techniques \vb{up to few eV of energy scale}. The theoretical models explaining the dynamics of charge excitations such as acoustic and optical plasmons in \textcolor{black}{hole- and electron-doped} cuprates should properly account for the \textit{\textbf{k}}-DAES~\cite{Mitrano2018, Husain2019, Hepting2018, Nag2020, Nag2024, Kirsty2023, Greco2016, Greco2019}. Furthermore, we demonstrate that RIXS can be used to extract the size of the \com{electronic energy} gap \com{$\Delta$}, making it a complementary method to other established probes such as ARPES and STM\com{, and opening the route to the ultrafast regime}. Our findings, in combination with high-energy resolution RIXS lay the foundation for studying the electron dynamics and the \com{energy} gap size in bulk and complex heterostructures, buried, and twisted layers~\cite{Zhu2021, Can2021, Ju2022, Zhao2023}. \textcolor{black}{For instance, the RIXS has unveiled the peculiar electron behaviors in superconducting infinite layer nickelates which however cannot be accessed by the surface-sensitive probes due to the SrTiO$_3$ capping layer~\cite{Li2019, Lu2021, Hepting2021, Fan2024}.}
    

    \begin{acknowledgments}
    Work performed at Brookhaven National Laboratory was supported by the U.S. Department of Energy (DOE), Division of Materials Science, under Contract No. DE-SC0012704. This research uses the beamline 2-ID of the National Synchrotron Light Source II, a DOE Office of Science User Facility operated for the DOE Office of Science by Brookhaven National Laboratory under Contract No. DE-SC0012704. This work was supported by JSPS KAKENHI Grant Number JP24K06943.
    \end{acknowledgments}


\begin{thebibliography}{61}%
\makeatletter
\providecommand \@ifxundefined [1]{%
 \@ifx{#1\undefined}
}%
\providecommand \@ifnum [1]{%
 \ifnum #1\expandafter \@firstoftwo
 \else \expandafter \@secondoftwo
 \fi
}%
\providecommand \@ifx [1]{%
 \ifx #1\expandafter \@firstoftwo
 \else \expandafter \@secondoftwo
 \fi
}%
\providecommand \natexlab [1]{#1}%
\providecommand \enquote  [1]{``#1''}%
\providecommand \bibnamefont  [1]{#1}%
\providecommand \bibfnamefont [1]{#1}%
\providecommand \citenamefont [1]{#1}%
\providecommand \href@noop [0]{\@secondoftwo}%
\providecommand \href [0]{\begingroup \@sanitize@url \@href}%
\providecommand \@href[1]{\@@startlink{#1}\@@href}%
\providecommand \@@href[1]{\endgroup#1\@@endlink}%
\providecommand \@sanitize@url [0]{\catcode `\\12\catcode `\$12\catcode `\&12\catcode `\#12\catcode `\^12\catcode `\_12\catcode `\%12\relax}%
\providecommand \@@startlink[1]{}%
\providecommand \@@endlink[0]{}%
\providecommand \url  [0]{\begingroup\@sanitize@url \@url }%
\providecommand \@url [1]{\endgroup\@href {#1}{\urlprefix }}%
\providecommand \urlprefix  [0]{URL }%
\providecommand \Eprint [0]{\href }%
\providecommand \doibase [0]{https://doi.org/}%
\providecommand \selectlanguage [0]{\@gobble}%
\providecommand \bibinfo  [0]{\@secondoftwo}%
\providecommand \bibfield  [0]{\@secondoftwo}%
\providecommand \translation [1]{[#1]}%
\providecommand \BibitemOpen [0]{}%
\providecommand \bibitemStop [0]{}%
\providecommand \bibitemNoStop [0]{.\EOS\space}%
\providecommand \EOS [0]{\spacefactor3000\relax}%
\providecommand \BibitemShut  [1]{\csname bibitem#1\endcsname}%
\let\auto@bib@innerbib\@empty
\bibitem [{\citenamefont {Bardeen}\ \emph {et~al.}(1957)\citenamefont {Bardeen}, \citenamefont {Cooper},\ and\ \citenamefont {Schrieffer}}]{Bardeen1957}%
  \BibitemOpen
  \bibfield  {author} {\bibinfo {author} {\bibfnamefont {J.}~\bibnamefont {Bardeen}}, \bibinfo {author} {\bibfnamefont {L.~N.}\ \bibnamefont {Cooper}},\ and\ \bibinfo {author} {\bibfnamefont {J.~R.}\ \bibnamefont {Schrieffer}},\ }\bibfield  {title} {\bibinfo {title} {Theory of superconductivity},\ }\href {https://doi.org/10.1103/PhysRev.108.1175} {\bibfield  {journal} {\bibinfo  {journal} {Phys. Rev.}\ }\textbf {\bibinfo {volume} {108}},\ \bibinfo {pages} {1175} (\bibinfo {year} {1957})}\BibitemShut {NoStop}%
\bibitem [{\citenamefont {Bednorz}\ and\ \citenamefont {Müller}(1986)}]{Bednorz1986}%
  \BibitemOpen
  \bibfield  {author} {\bibinfo {author} {\bibfnamefont {J.~G.}\ \bibnamefont {Bednorz}}\ and\ \bibinfo {author} {\bibfnamefont {K.~A.}\ \bibnamefont {Müller}},\ }\bibfield  {title} {\bibinfo {title} {{Possible high $T_c$ superconductivity in the Ba-La-Cu-O system}},\ }\href {https://doi.org/10.1007/BF01303701} {\bibfield  {journal} {\bibinfo  {journal} {Zeitschrift für Physik B Condensed Matter}\ }\textbf {\bibinfo {volume} {64}},\ \bibinfo {pages} {189} (\bibinfo {year} {1986})}\BibitemShut {NoStop}%
\bibitem [{\citenamefont {Nagamatsu}\ \emph {et~al.}(2001)\citenamefont {Nagamatsu}, \citenamefont {Nakagawa}, \citenamefont {Muranaka}, \citenamefont {Zenitani},\ and\ \citenamefont {Akimitsu}}]{Nagamatsu2001}%
  \BibitemOpen
  \bibfield  {author} {\bibinfo {author} {\bibfnamefont {J.}~\bibnamefont {Nagamatsu}}, \bibinfo {author} {\bibfnamefont {N.}~\bibnamefont {Nakagawa}}, \bibinfo {author} {\bibfnamefont {T.}~\bibnamefont {Muranaka}}, \bibinfo {author} {\bibfnamefont {Y.}~\bibnamefont {Zenitani}},\ and\ \bibinfo {author} {\bibfnamefont {J.}~\bibnamefont {Akimitsu}},\ }\bibfield  {title} {\bibinfo {title} {Superconductivity at 39k in magnesium diboride},\ }\href {https://doi.org/10.1038/35065039} {\bibfield  {journal} {\bibinfo  {journal} {Nature}\ }\textbf {\bibinfo {volume} {410}},\ \bibinfo {pages} {63} (\bibinfo {year} {2001})}\BibitemShut {NoStop}%
\bibitem [{\citenamefont {Keimer}\ \emph {et~al.}(2015)\citenamefont {Keimer}, \citenamefont {Kivelson}, \citenamefont {Norman}, \citenamefont {Uchida},\ and\ \citenamefont {Zaanen}}]{Keimer2015}%
  \BibitemOpen
  \bibfield  {author} {\bibinfo {author} {\bibfnamefont {B.}~\bibnamefont {Keimer}}, \bibinfo {author} {\bibfnamefont {S.~A.}\ \bibnamefont {Kivelson}}, \bibinfo {author} {\bibfnamefont {M.~R.}\ \bibnamefont {Norman}}, \bibinfo {author} {\bibfnamefont {S.}~\bibnamefont {Uchida}},\ and\ \bibinfo {author} {\bibfnamefont {J.}~\bibnamefont {Zaanen}},\ }\bibfield  {title} {\bibinfo {title} {From quantum matter to high-temperature superconductivity in copper oxides},\ }\href {https://doi.org/10.1038/nature14165} {\bibfield  {journal} {\bibinfo  {journal} {Nature}\ }\textbf {\bibinfo {volume} {518}},\ \bibinfo {pages} {179} (\bibinfo {year} {2015})}\BibitemShut {NoStop}%
\bibitem [{\citenamefont {Thomas}\ \emph {et~al.}(1988)\citenamefont {Thomas}, \citenamefont {Orenstein}, \citenamefont {Rapkine}, \citenamefont {Capizzi}, \citenamefont {Millis}, \citenamefont {Bhatt}, \citenamefont {Schneemeyer},\ and\ \citenamefont {Waszczak}}]{Thomas1988}%
  \BibitemOpen
  \bibfield  {author} {\bibinfo {author} {\bibfnamefont {G.~A.}\ \bibnamefont {Thomas}}, \bibinfo {author} {\bibfnamefont {J.}~\bibnamefont {Orenstein}}, \bibinfo {author} {\bibfnamefont {D.~H.}\ \bibnamefont {Rapkine}}, \bibinfo {author} {\bibfnamefont {M.}~\bibnamefont {Capizzi}}, \bibinfo {author} {\bibfnamefont {A.~J.}\ \bibnamefont {Millis}}, \bibinfo {author} {\bibfnamefont {R.~N.}\ \bibnamefont {Bhatt}}, \bibinfo {author} {\bibfnamefont {L.~F.}\ \bibnamefont {Schneemeyer}},\ and\ \bibinfo {author} {\bibfnamefont {J.~V.}\ \bibnamefont {Waszczak}},\ }\bibfield  {title} {\bibinfo {title} {{Ba$_2$YCu$_3$O$_{7-\delta}$: Electrodynamics of Crystals with High Reflectivity}},\ }\href {https://doi.org/10.1103/PhysRevLett.61.1313} {\bibfield  {journal} {\bibinfo  {journal} {Phys. Rev. Lett.}\ }\textbf {\bibinfo {volume} {61}},\ \bibinfo {pages} {1313} (\bibinfo {year} {1988})}\BibitemShut {NoStop}%
\bibitem [{\citenamefont {Collins}\ \emph {et~al.}(1989)\citenamefont {Collins}, \citenamefont {Schlesinger}, \citenamefont {Holtzberg}, \citenamefont {Chaudhari},\ and\ \citenamefont {Feild}}]{Collins1989}%
  \BibitemOpen
  \bibfield  {author} {\bibinfo {author} {\bibfnamefont {R.~T.}\ \bibnamefont {Collins}}, \bibinfo {author} {\bibfnamefont {Z.}~\bibnamefont {Schlesinger}}, \bibinfo {author} {\bibfnamefont {F.}~\bibnamefont {Holtzberg}}, \bibinfo {author} {\bibfnamefont {P.}~\bibnamefont {Chaudhari}},\ and\ \bibinfo {author} {\bibfnamefont {C.}~\bibnamefont {Feild}},\ }\bibfield  {title} {\bibinfo {title} {{Reflectivity and conductivity of ${\mathrm{YBa}}_{2}$${\mathrm{Cu}}_{3}$${\mathrm{O}}_{7}$}},\ }\href {https://doi.org/10.1103/PhysRevB.39.6571} {\bibfield  {journal} {\bibinfo  {journal} {Phys. Rev. B}\ }\textbf {\bibinfo {volume} {39}},\ \bibinfo {pages} {6571} (\bibinfo {year} {1989})}\BibitemShut {NoStop}%
\bibitem [{\citenamefont {Bo\v{z}ovi\'{c}}\ \emph {et~al.}(1991)\citenamefont {Bo\v{z}ovi\'{c}}, \citenamefont {Kim}, \citenamefont {Harris},\ and\ \citenamefont {Lee}}]{Ivan1991}%
  \BibitemOpen
  \bibfield  {author} {\bibinfo {author} {\bibfnamefont {I.}~\bibnamefont {Bo\v{z}ovi\'{c}}}, \bibinfo {author} {\bibfnamefont {J.~H.}\ \bibnamefont {Kim}}, \bibinfo {author} {\bibfnamefont {J.~S.}\ \bibnamefont {Harris}},\ and\ \bibinfo {author} {\bibfnamefont {W.~Y.}\ \bibnamefont {Lee}},\ }\bibfield  {title} {\bibinfo {title} {{Optical study of plasmons in ${\mathrm{Tl}}_{2}$${\mathrm{Ba}}_{2}$${\mathrm{Ca}}_{2}$${\mathrm{Cu}}_{3}$${\mathrm{O}}_{10}$}},\ }\href {https://doi.org/10.1103/PhysRevB.43.1169} {\bibfield  {journal} {\bibinfo  {journal} {Phys. Rev. B}\ }\textbf {\bibinfo {volume} {43}},\ \bibinfo {pages} {1169} (\bibinfo {year} {1991})}\BibitemShut {NoStop}%
\bibitem [{\citenamefont {Cooper}\ \emph {et~al.}(1992)\citenamefont {Cooper}, \citenamefont {Kotz}, \citenamefont {Karlow}, \citenamefont {Klein}, \citenamefont {Lee}, \citenamefont {Giapintzakis},\ and\ \citenamefont {Ginsberg}}]{Cooper1992}%
  \BibitemOpen
  \bibfield  {author} {\bibinfo {author} {\bibfnamefont {S.~L.}\ \bibnamefont {Cooper}}, \bibinfo {author} {\bibfnamefont {A.~L.}\ \bibnamefont {Kotz}}, \bibinfo {author} {\bibfnamefont {M.~A.}\ \bibnamefont {Karlow}}, \bibinfo {author} {\bibfnamefont {M.~V.}\ \bibnamefont {Klein}}, \bibinfo {author} {\bibfnamefont {W.~C.}\ \bibnamefont {Lee}}, \bibinfo {author} {\bibfnamefont {J.}~\bibnamefont {Giapintzakis}},\ and\ \bibinfo {author} {\bibfnamefont {D.~M.}\ \bibnamefont {Ginsberg}},\ }\bibfield  {title} {\bibinfo {title} {{Development of the optical conductivity with doping in single-domain ${\mathrm{YBa}}_{2}$${\mathrm{Cu}}_{3}$${\mathrm{O}}_{6+\mathit{x}}$}},\ }\href {https://doi.org/10.1103/PhysRevB.45.2549} {\bibfield  {journal} {\bibinfo  {journal} {Phys. Rev. B}\ }\textbf {\bibinfo {volume} {45}},\ \bibinfo {pages} {2549} (\bibinfo {year} {1992})}\BibitemShut {NoStop}%
\bibitem [{\citenamefont {Abrahams}\ and\ \citenamefont {Varma}(2000)}]{Elihu2000}%
  \BibitemOpen
  \bibfield  {author} {\bibinfo {author} {\bibfnamefont {E.}~\bibnamefont {Abrahams}}\ and\ \bibinfo {author} {\bibfnamefont {C.~M.}\ \bibnamefont {Varma}},\ }\bibfield  {title} {\bibinfo {title} {What angle-resolved photoemission experiments tell about the microscopic theory for high-temperature superconductors},\ }\href {https://doi.org/10.1073/pnas.100118797} {\bibfield  {journal} {\bibinfo  {journal} {Proceedings of the National Academy of Sciences}\ }\textbf {\bibinfo {volume} {97}},\ \bibinfo {pages} {5714} (\bibinfo {year} {2000})}\BibitemShut {NoStop}%
\bibitem [{\citenamefont {Bonn}\ \emph {et~al.}(1992)\citenamefont {Bonn}, \citenamefont {Dosanjh}, \citenamefont {Liang},\ and\ \citenamefont {Hardy}}]{Bonn1992}%
  \BibitemOpen
  \bibfield  {author} {\bibinfo {author} {\bibfnamefont {D.~A.}\ \bibnamefont {Bonn}}, \bibinfo {author} {\bibfnamefont {P.}~\bibnamefont {Dosanjh}}, \bibinfo {author} {\bibfnamefont {R.}~\bibnamefont {Liang}},\ and\ \bibinfo {author} {\bibfnamefont {W.~N.}\ \bibnamefont {Hardy}},\ }\bibfield  {title} {\bibinfo {title} {{Evidence for rapid suppression of quasiparticle scattering below ${\mathit{T}}_{\mathit{c}}$ in ${\mathrm{YBa}}_{2}$${\mathrm{Cu}}_{3}$${\mathrm{O}}_{7\mathrm{\ensuremath{-}}\mathrm{\ensuremath{\delta}}}$}},\ }\href {https://doi.org/10.1103/PhysRevLett.68.2390} {\bibfield  {journal} {\bibinfo  {journal} {Phys. Rev. Lett.}\ }\textbf {\bibinfo {volume} {68}},\ \bibinfo {pages} {2390} (\bibinfo {year} {1992})}\BibitemShut {NoStop}%
\bibitem [{\citenamefont {Rieck}\ \emph {et~al.}(1995)\citenamefont {Rieck}, \citenamefont {Little}, \citenamefont {Ruvalds},\ and\ \citenamefont {Virosztek}}]{Rieck1995}%
  \BibitemOpen
  \bibfield  {author} {\bibinfo {author} {\bibfnamefont {C.~T.}\ \bibnamefont {Rieck}}, \bibinfo {author} {\bibfnamefont {W.~A.}\ \bibnamefont {Little}}, \bibinfo {author} {\bibfnamefont {J.}~\bibnamefont {Ruvalds}},\ and\ \bibinfo {author} {\bibfnamefont {A.}~\bibnamefont {Virosztek}},\ }\bibfield  {title} {\bibinfo {title} {Infrared and microwave spectra of an energy gap in high-temperature superconductors},\ }\href {https://doi.org/10.1103/PhysRevB.51.3772} {\bibfield  {journal} {\bibinfo  {journal} {Phys. Rev. B}\ }\textbf {\bibinfo {volume} {51}},\ \bibinfo {pages} {3772} (\bibinfo {year} {1995})}\BibitemShut {NoStop}%
\bibitem [{\citenamefont {Varma}\ \emph {et~al.}(1989)\citenamefont {Varma}, \citenamefont {Littlewood}, \citenamefont {Schmitt-Rink}, \citenamefont {Abrahams},\ and\ \citenamefont {Ruckenstein}}]{Varma1989}%
  \BibitemOpen
  \bibfield  {author} {\bibinfo {author} {\bibfnamefont {C.~M.}\ \bibnamefont {Varma}}, \bibinfo {author} {\bibfnamefont {P.~B.}\ \bibnamefont {Littlewood}}, \bibinfo {author} {\bibfnamefont {S.}~\bibnamefont {Schmitt-Rink}}, \bibinfo {author} {\bibfnamefont {E.}~\bibnamefont {Abrahams}},\ and\ \bibinfo {author} {\bibfnamefont {A.~E.}\ \bibnamefont {Ruckenstein}},\ }\bibfield  {title} {\bibinfo {title} {{Phenomenology of the normal state of Cu-O high-temperature superconductors}},\ }\href {https://doi.org/10.1103/PhysRevLett.63.1996} {\bibfield  {journal} {\bibinfo  {journal} {Phys. Rev. Lett.}\ }\textbf {\bibinfo {volume} {63}},\ \bibinfo {pages} {1996} (\bibinfo {year} {1989})}\BibitemShut {NoStop}%
\bibitem [{\citenamefont {Abdel-Jawad}\ \emph {et~al.}(2006)\citenamefont {Abdel-Jawad}, \citenamefont {Kennett}, \citenamefont {Balicas}, \citenamefont {Carrington}, \citenamefont {Mackenzie}, \citenamefont {McKenzie},\ and\ \citenamefont {Hussey}}]{Abdel2006}%
  \BibitemOpen
  \bibfield  {author} {\bibinfo {author} {\bibfnamefont {M.}~\bibnamefont {Abdel-Jawad}}, \bibinfo {author} {\bibfnamefont {M.~P.}\ \bibnamefont {Kennett}}, \bibinfo {author} {\bibfnamefont {L.}~\bibnamefont {Balicas}}, \bibinfo {author} {\bibfnamefont {A.}~\bibnamefont {Carrington}}, \bibinfo {author} {\bibfnamefont {A.~P.}\ \bibnamefont {Mackenzie}}, \bibinfo {author} {\bibfnamefont {R.~H.}\ \bibnamefont {McKenzie}},\ and\ \bibinfo {author} {\bibfnamefont {N.~E.}\ \bibnamefont {Hussey}},\ }\bibfield  {title} {\bibinfo {title} {Anisotropic scattering and anomalous normal-state transport in a high-temperature superconductor},\ }\href {https://doi.org/10.1038/nphys449} {\bibfield  {journal} {\bibinfo  {journal} {Nature Physics}\ }\textbf {\bibinfo {volume} {2}},\ \bibinfo {pages} {821} (\bibinfo {year} {2006})}\BibitemShut {NoStop}%
\bibitem [{\citenamefont {Fang}\ \emph {et~al.}(2022)\citenamefont {Fang}, \citenamefont {Grissonnanche}, \citenamefont {Legros}, \citenamefont {Verret}, \citenamefont {Laliberté}, \citenamefont {Collignon}, \citenamefont {Ataei}, \citenamefont {Dion}, \citenamefont {Zhou}, \citenamefont {Graf}, \citenamefont {Lawler}, \citenamefont {Goddard}, \citenamefont {Taillefer},\ and\ \citenamefont {Ramshaw}}]{Fang2022}%
  \BibitemOpen
  \bibfield  {author} {\bibinfo {author} {\bibfnamefont {Y.}~\bibnamefont {Fang}}, \bibinfo {author} {\bibfnamefont {G.}~\bibnamefont {Grissonnanche}}, \bibinfo {author} {\bibfnamefont {A.}~\bibnamefont {Legros}}, \bibinfo {author} {\bibfnamefont {S.}~\bibnamefont {Verret}}, \bibinfo {author} {\bibfnamefont {F.}~\bibnamefont {Laliberté}}, \bibinfo {author} {\bibfnamefont {C.}~\bibnamefont {Collignon}}, \bibinfo {author} {\bibfnamefont {A.}~\bibnamefont {Ataei}}, \bibinfo {author} {\bibfnamefont {M.}~\bibnamefont {Dion}}, \bibinfo {author} {\bibfnamefont {J.}~\bibnamefont {Zhou}}, \bibinfo {author} {\bibfnamefont {D.}~\bibnamefont {Graf}}, \bibinfo {author} {\bibfnamefont {M.~J.}\ \bibnamefont {Lawler}}, \bibinfo {author} {\bibfnamefont {P.~A.}\ \bibnamefont {Goddard}}, \bibinfo {author} {\bibfnamefont {L.}~\bibnamefont {Taillefer}},\ and\ \bibinfo {author} {\bibfnamefont {B.~J.}\ \bibnamefont {Ramshaw}},\ }\bibfield  {title} {\bibinfo {title} {Fermi surface transformation at the pseudogap critical point of a
  cuprate superconductor},\ }\href {https://doi.org/10.1038/s41567-022-01514-1} {\bibfield  {journal} {\bibinfo  {journal} {Nature Physics}\ }\textbf {\bibinfo {volume} {18}},\ \bibinfo {pages} {558} (\bibinfo {year} {2022})}\BibitemShut {NoStop}%
\bibitem [{\citenamefont {Valla}\ \emph {et~al.}(1999)\citenamefont {Valla}, \citenamefont {Fedorov}, \citenamefont {Johnson}, \citenamefont {Wells}, \citenamefont {Hulbert}, \citenamefont {Li}, \citenamefont {Gu},\ and\ \citenamefont {Koshizuka}}]{Valla1999}%
  \BibitemOpen
  \bibfield  {author} {\bibinfo {author} {\bibfnamefont {T.}~\bibnamefont {Valla}}, \bibinfo {author} {\bibfnamefont {A.~V.}\ \bibnamefont {Fedorov}}, \bibinfo {author} {\bibfnamefont {P.~D.}\ \bibnamefont {Johnson}}, \bibinfo {author} {\bibfnamefont {B.~O.}\ \bibnamefont {Wells}}, \bibinfo {author} {\bibfnamefont {S.~L.}\ \bibnamefont {Hulbert}}, \bibinfo {author} {\bibfnamefont {Q.}~\bibnamefont {Li}}, \bibinfo {author} {\bibfnamefont {G.~D.}\ \bibnamefont {Gu}},\ and\ \bibinfo {author} {\bibfnamefont {N.}~\bibnamefont {Koshizuka}},\ }\bibfield  {title} {\bibinfo {title} {{Evidence for Quantum Critical Behavior in the Optimally Doped Cuprate Bi$_2$Sr$_2$CaCu$_2$O$_{8+\delta}$}},\ }\href {https://doi.org/10.1126/science.285.5436.2110} {\bibfield  {journal} {\bibinfo  {journal} {Science}\ }\textbf {\bibinfo {volume} {285}},\ \bibinfo {pages} {2110} (\bibinfo {year} {1999})}\BibitemShut {NoStop}%
\bibitem [{\citenamefont {Valla}\ \emph {et~al.}(2000)\citenamefont {Valla}, \citenamefont {Fedorov}, \citenamefont {Johnson}, \citenamefont {Li}, \citenamefont {Gu},\ and\ \citenamefont {Koshizuka}}]{Valla2000}%
  \BibitemOpen
  \bibfield  {author} {\bibinfo {author} {\bibfnamefont {T.}~\bibnamefont {Valla}}, \bibinfo {author} {\bibfnamefont {A.~V.}\ \bibnamefont {Fedorov}}, \bibinfo {author} {\bibfnamefont {P.~D.}\ \bibnamefont {Johnson}}, \bibinfo {author} {\bibfnamefont {Q.}~\bibnamefont {Li}}, \bibinfo {author} {\bibfnamefont {G.~D.}\ \bibnamefont {Gu}},\ and\ \bibinfo {author} {\bibfnamefont {N.}~\bibnamefont {Koshizuka}},\ }\bibfield  {title} {\bibinfo {title} {{Temperature Dependent Scattering Rates at the Fermi Surface of Optimally Doped ${\mathrm{Bi}}_{2}{\mathrm{Sr}}_{2}{\mathrm{CaCu}}_{2}{O}_{8+\mathit{\ensuremath{\delta}}}$}},\ }\href {https://doi.org/10.1103/PhysRevLett.85.828} {\bibfield  {journal} {\bibinfo  {journal} {Phys. Rev. Lett.}\ }\textbf {\bibinfo {volume} {85}},\ \bibinfo {pages} {828} (\bibinfo {year} {2000})}\BibitemShut {NoStop}%
\bibitem [{\citenamefont {Kaminski}\ \emph {et~al.}(2005)\citenamefont {Kaminski}, \citenamefont {Fretwell}, \citenamefont {Norman}, \citenamefont {Randeria}, \citenamefont {Rosenkranz}, \citenamefont {Chatterjee}, \citenamefont {Campuzano}, \citenamefont {Mesot}, \citenamefont {Sato}, \citenamefont {Takahashi}, \citenamefont {Terashima}, \citenamefont {Takano}, \citenamefont {Kadowaki}, \citenamefont {Li},\ and\ \citenamefont {Raffy}}]{Kaminski2005}%
  \BibitemOpen
  \bibfield  {author} {\bibinfo {author} {\bibfnamefont {A.}~\bibnamefont {Kaminski}}, \bibinfo {author} {\bibfnamefont {H.~M.}\ \bibnamefont {Fretwell}}, \bibinfo {author} {\bibfnamefont {M.~R.}\ \bibnamefont {Norman}}, \bibinfo {author} {\bibfnamefont {M.}~\bibnamefont {Randeria}}, \bibinfo {author} {\bibfnamefont {S.}~\bibnamefont {Rosenkranz}}, \bibinfo {author} {\bibfnamefont {U.}~\bibnamefont {Chatterjee}}, \bibinfo {author} {\bibfnamefont {J.~C.}\ \bibnamefont {Campuzano}}, \bibinfo {author} {\bibfnamefont {J.}~\bibnamefont {Mesot}}, \bibinfo {author} {\bibfnamefont {T.}~\bibnamefont {Sato}}, \bibinfo {author} {\bibfnamefont {T.}~\bibnamefont {Takahashi}}, \bibinfo {author} {\bibfnamefont {T.}~\bibnamefont {Terashima}}, \bibinfo {author} {\bibfnamefont {M.}~\bibnamefont {Takano}}, \bibinfo {author} {\bibfnamefont {K.}~\bibnamefont {Kadowaki}}, \bibinfo {author} {\bibfnamefont {Z.~Z.}\ \bibnamefont {Li}},\ and\ \bibinfo {author} {\bibfnamefont {H.}~\bibnamefont {Raffy}},\ }\bibfield  {title} {\bibinfo
  {title} {Momentum anisotropy of the scattering rate in cuprate superconductors},\ }\href {https://doi.org/10.1103/PhysRevB.71.014517} {\bibfield  {journal} {\bibinfo  {journal} {Phys. Rev. B}\ }\textbf {\bibinfo {volume} {71}},\ \bibinfo {pages} {014517} (\bibinfo {year} {2005})}\BibitemShut {NoStop}%
\bibitem [{\citenamefont {Chang}\ \emph {et~al.}(2013)\citenamefont {Chang}, \citenamefont {Mansson}, \citenamefont {Pailhes}, \citenamefont {Claesson}, \citenamefont {Lipscombe}, \citenamefont {Hayden}, \citenamefont {Patthey}, \citenamefont {Tjernberg},\ and\ \citenamefont {Mesot}}]{Chang2013}%
  \BibitemOpen
  \bibfield  {author} {\bibinfo {author} {\bibfnamefont {J.}~\bibnamefont {Chang}}, \bibinfo {author} {\bibfnamefont {M.}~\bibnamefont {Mansson}}, \bibinfo {author} {\bibfnamefont {S.}~\bibnamefont {Pailhes}}, \bibinfo {author} {\bibfnamefont {T.}~\bibnamefont {Claesson}}, \bibinfo {author} {\bibfnamefont {O.~J.}\ \bibnamefont {Lipscombe}}, \bibinfo {author} {\bibfnamefont {S.~M.}\ \bibnamefont {Hayden}}, \bibinfo {author} {\bibfnamefont {L.}~\bibnamefont {Patthey}}, \bibinfo {author} {\bibfnamefont {O.}~\bibnamefont {Tjernberg}},\ and\ \bibinfo {author} {\bibfnamefont {J.}~\bibnamefont {Mesot}},\ }\bibfield  {title} {\bibinfo {title} {{Anisotropic breakdown of Fermi liquid quasiparticle excitations in overdoped La$_{2-x}$Sr$_x$CuO$_4$}},\ }\href {https://doi.org/10.1038/ncomms3559} {\bibfield  {journal} {\bibinfo  {journal} {Nature Communications}\ }\textbf {\bibinfo {volume} {4}},\ \bibinfo {pages} {2559} (\bibinfo {year} {2013})}\BibitemShut {NoStop}%
\bibitem [{\citenamefont {Varma}\ and\ \citenamefont {Abrahams}(2001)}]{Varma2001}%
  \BibitemOpen
  \bibfield  {author} {\bibinfo {author} {\bibfnamefont {C.~M.}\ \bibnamefont {Varma}}\ and\ \bibinfo {author} {\bibfnamefont {E.}~\bibnamefont {Abrahams}},\ }\bibfield  {title} {\bibinfo {title} {{Effective Lorentz Force due to Small-Angle Impurity Scattering: Magnetotransport in High- ${T}_{c}$ Superconductors}},\ }\href {https://doi.org/10.1103/PhysRevLett.86.4652} {\bibfield  {journal} {\bibinfo  {journal} {Phys. Rev. Lett.}\ }\textbf {\bibinfo {volume} {86}},\ \bibinfo {pages} {4652} (\bibinfo {year} {2001})}\BibitemShut {NoStop}%
\bibitem [{\citenamefont {Norman}(1990)}]{Norman1990}%
  \BibitemOpen
  \bibfield  {author} {\bibinfo {author} {\bibfnamefont {M.~R.}\ \bibnamefont {Norman}},\ }\bibfield  {title} {\bibinfo {title} {{Anisotropic exchange and superconductivity in ${\mathrm{UPt}}_{3}$}},\ }\href {https://doi.org/10.1103/PhysRevB.41.170} {\bibfield  {journal} {\bibinfo  {journal} {Phys. Rev. B}\ }\textbf {\bibinfo {volume} {41}},\ \bibinfo {pages} {170} (\bibinfo {year} {1990})}\BibitemShut {NoStop}%
\bibitem [{\citenamefont {Millis}(1992)}]{Millis1992}%
  \BibitemOpen
  \bibfield  {author} {\bibinfo {author} {\bibfnamefont {A.~J.}\ \bibnamefont {Millis}},\ }\bibfield  {title} {\bibinfo {title} {Nearly antiferromagnetic fermi liquids: An analytic eliashberg approach},\ }\href {https://doi.org/10.1103/PhysRevB.45.13047} {\bibfield  {journal} {\bibinfo  {journal} {Phys. Rev. B}\ }\textbf {\bibinfo {volume} {45}},\ \bibinfo {pages} {13047} (\bibinfo {year} {1992})}\BibitemShut {NoStop}%
\bibitem [{\citenamefont {Dahm}\ \emph {et~al.}(2005)\citenamefont {Dahm}, \citenamefont {Hirschfeld}, \citenamefont {Scalapino},\ and\ \citenamefont {Zhu}}]{Dahm2005}%
  \BibitemOpen
  \bibfield  {author} {\bibinfo {author} {\bibfnamefont {T.}~\bibnamefont {Dahm}}, \bibinfo {author} {\bibfnamefont {P.~J.}\ \bibnamefont {Hirschfeld}}, \bibinfo {author} {\bibfnamefont {D.~J.}\ \bibnamefont {Scalapino}},\ and\ \bibinfo {author} {\bibfnamefont {L.}~\bibnamefont {Zhu}},\ }\bibfield  {title} {\bibinfo {title} {Nodal quasiparticle lifetimes in cuprate superconductors},\ }\href {https://doi.org/10.1103/PhysRevB.72.214512} {\bibfield  {journal} {\bibinfo  {journal} {Phys. Rev. B}\ }\textbf {\bibinfo {volume} {72}},\ \bibinfo {pages} {214512} (\bibinfo {year} {2005})}\BibitemShut {NoStop}%
\bibitem [{\citenamefont {Mitrano}\ \emph {et~al.}(2018)\citenamefont {Mitrano}, \citenamefont {Husain}, \citenamefont {Vig}, \citenamefont {Kogar}, \citenamefont {Rak}, \citenamefont {Rubeck}, \citenamefont {Schmalian}, \citenamefont {Uchoa}, \citenamefont {Schneeloch}, \citenamefont {Zhong}, \citenamefont {Gu},\ and\ \citenamefont {Abbamonte}}]{Mitrano2018}%
  \BibitemOpen
  \bibfield  {author} {\bibinfo {author} {\bibfnamefont {M.}~\bibnamefont {Mitrano}}, \bibinfo {author} {\bibfnamefont {A.~A.}\ \bibnamefont {Husain}}, \bibinfo {author} {\bibfnamefont {S.}~\bibnamefont {Vig}}, \bibinfo {author} {\bibfnamefont {A.}~\bibnamefont {Kogar}}, \bibinfo {author} {\bibfnamefont {M.~S.}\ \bibnamefont {Rak}}, \bibinfo {author} {\bibfnamefont {S.~I.}\ \bibnamefont {Rubeck}}, \bibinfo {author} {\bibfnamefont {J.}~\bibnamefont {Schmalian}}, \bibinfo {author} {\bibfnamefont {B.}~\bibnamefont {Uchoa}}, \bibinfo {author} {\bibfnamefont {J.}~\bibnamefont {Schneeloch}}, \bibinfo {author} {\bibfnamefont {R.}~\bibnamefont {Zhong}}, \bibinfo {author} {\bibfnamefont {G.~D.}\ \bibnamefont {Gu}},\ and\ \bibinfo {author} {\bibfnamefont {P.}~\bibnamefont {Abbamonte}},\ }\bibfield  {title} {\bibinfo {title} {Anomalous density fluctuations in a strange metal},\ }\href {https://doi.org/10.1073/pnas.1721495115} {\bibfield  {journal} {\bibinfo  {journal} {Proceedings of the National Academy of Sciences}\
  }\textbf {\bibinfo {volume} {115}},\ \bibinfo {pages} {5392} (\bibinfo {year} {2018})}\BibitemShut {NoStop}%
\bibitem [{\citenamefont {Husain}\ \emph {et~al.}(2019)\citenamefont {Husain}, \citenamefont {Mitrano}, \citenamefont {Rak}, \citenamefont {Rubeck}, \citenamefont {Uchoa}, \citenamefont {March}, \citenamefont {Dwyer}, \citenamefont {Schneeloch}, \citenamefont {Zhong}, \citenamefont {Gu},\ and\ \citenamefont {Abbamonte}}]{Husain2019}%
  \BibitemOpen
  \bibfield  {author} {\bibinfo {author} {\bibfnamefont {A.~A.}\ \bibnamefont {Husain}}, \bibinfo {author} {\bibfnamefont {M.}~\bibnamefont {Mitrano}}, \bibinfo {author} {\bibfnamefont {M.~S.}\ \bibnamefont {Rak}}, \bibinfo {author} {\bibfnamefont {S.}~\bibnamefont {Rubeck}}, \bibinfo {author} {\bibfnamefont {B.}~\bibnamefont {Uchoa}}, \bibinfo {author} {\bibfnamefont {K.}~\bibnamefont {March}}, \bibinfo {author} {\bibfnamefont {C.}~\bibnamefont {Dwyer}}, \bibinfo {author} {\bibfnamefont {J.}~\bibnamefont {Schneeloch}}, \bibinfo {author} {\bibfnamefont {R.}~\bibnamefont {Zhong}}, \bibinfo {author} {\bibfnamefont {G.~D.}\ \bibnamefont {Gu}},\ and\ \bibinfo {author} {\bibfnamefont {P.}~\bibnamefont {Abbamonte}},\ }\bibfield  {title} {\bibinfo {title} {Crossover of charge fluctuations across the strange metal phase diagram},\ }\href {https://doi.org/10.1103/PhysRevX.9.041062} {\bibfield  {journal} {\bibinfo  {journal} {Phys. Rev. X}\ }\textbf {\bibinfo {volume} {9}},\ \bibinfo {pages} {041062} (\bibinfo {year}
  {2019})}\BibitemShut {NoStop}%
\bibitem [{\citenamefont {Ament}\ \emph {et~al.}(2011)\citenamefont {Ament}, \citenamefont {van Veenendaal}, \citenamefont {Devereaux}, \citenamefont {Hill},\ and\ \citenamefont {van~den Brink}}]{Ament2011}%
  \BibitemOpen
  \bibfield  {author} {\bibinfo {author} {\bibfnamefont {L.~J.~P.}\ \bibnamefont {Ament}}, \bibinfo {author} {\bibfnamefont {M.}~\bibnamefont {van Veenendaal}}, \bibinfo {author} {\bibfnamefont {T.~P.}\ \bibnamefont {Devereaux}}, \bibinfo {author} {\bibfnamefont {J.~P.}\ \bibnamefont {Hill}},\ and\ \bibinfo {author} {\bibfnamefont {J.}~\bibnamefont {van~den Brink}},\ }\bibfield  {title} {\bibinfo {title} {Resonant inelastic x-ray scattering studies of elementary excitations},\ }\href {https://doi.org/10.1103/RevModPhys.83.705} {\bibfield  {journal} {\bibinfo  {journal} {Rev. Mod. Phys.}\ }\textbf {\bibinfo {volume} {83}},\ \bibinfo {pages} {705} (\bibinfo {year} {2011})}\BibitemShut {NoStop}%
\bibitem [{\citenamefont {Mitrano}\ \emph {et~al.}(2024)\citenamefont {Mitrano}, \citenamefont {Johnston}, \citenamefont {Kim},\ and\ \citenamefont {Dean}}]{Mitrano2024exploring}%
  \BibitemOpen
  \bibfield  {author} {\bibinfo {author} {\bibfnamefont {M.}~\bibnamefont {Mitrano}}, \bibinfo {author} {\bibfnamefont {S.}~\bibnamefont {Johnston}}, \bibinfo {author} {\bibfnamefont {Y.-J.}\ \bibnamefont {Kim}},\ and\ \bibinfo {author} {\bibfnamefont {M.~P.~M.}\ \bibnamefont {Dean}},\ }\bibfield  {title} {\bibinfo {title} {Exploring quantum materials with resonant inelastic x-ray scattering},\ }\href {https://doi.org/10.1103/PhysRevX.14.040501} {\bibfield  {journal} {\bibinfo  {journal} {Phys. Rev. X}\ }\textbf {\bibinfo {volume} {14}},\ \bibinfo {pages} {040501} (\bibinfo {year} {2024})}\BibitemShut {NoStop}%
\bibitem [{\citenamefont {Marra}\ \emph {et~al.}(2013)\citenamefont {Marra}, \citenamefont {Sykora}, \citenamefont {Wohlfeld},\ and\ \citenamefont {van~den Brink}}]{Marra2013}%
  \BibitemOpen
  \bibfield  {author} {\bibinfo {author} {\bibfnamefont {P.}~\bibnamefont {Marra}}, \bibinfo {author} {\bibfnamefont {S.}~\bibnamefont {Sykora}}, \bibinfo {author} {\bibfnamefont {K.}~\bibnamefont {Wohlfeld}},\ and\ \bibinfo {author} {\bibfnamefont {J.}~\bibnamefont {van~den Brink}},\ }\bibfield  {title} {\bibinfo {title} {Resonant inelastic x-ray scattering as a probe of the phase and excitations of the order parameter of superconductors},\ }\href {https://doi.org/10.1103/PhysRevLett.110.117005} {\bibfield  {journal} {\bibinfo  {journal} {Phys. Rev. Lett.}\ }\textbf {\bibinfo {volume} {110}},\ \bibinfo {pages} {117005} (\bibinfo {year} {2013})}\BibitemShut {NoStop}%
\bibitem [{\citenamefont {Suzuki}\ \emph {et~al.}(2018)\citenamefont {Suzuki}, \citenamefont {Minola}, \citenamefont {Lu}, \citenamefont {Peng}, \citenamefont {Fumagalli}, \citenamefont {Lefrançois}, \citenamefont {Loew}, \citenamefont {Porras}, \citenamefont {Kummer}, \citenamefont {Betto}, \citenamefont {Ishida}, \citenamefont {Eisaki}, \citenamefont {Hu}, \citenamefont {Zhou}, \citenamefont {Haverkort}, \citenamefont {Brookes}, \citenamefont {Braicovich}, \citenamefont {Ghiringhelli}, \citenamefont {Le~Tacon},\ and\ \citenamefont {Keimer}}]{Suzuki2018}%
  \BibitemOpen
  \bibfield  {author} {\bibinfo {author} {\bibfnamefont {H.}~\bibnamefont {Suzuki}}, \bibinfo {author} {\bibfnamefont {M.}~\bibnamefont {Minola}}, \bibinfo {author} {\bibfnamefont {Y.}~\bibnamefont {Lu}}, \bibinfo {author} {\bibfnamefont {Y.}~\bibnamefont {Peng}}, \bibinfo {author} {\bibfnamefont {R.}~\bibnamefont {Fumagalli}}, \bibinfo {author} {\bibfnamefont {E.}~\bibnamefont {Lefrançois}}, \bibinfo {author} {\bibfnamefont {T.}~\bibnamefont {Loew}}, \bibinfo {author} {\bibfnamefont {J.}~\bibnamefont {Porras}}, \bibinfo {author} {\bibfnamefont {K.}~\bibnamefont {Kummer}}, \bibinfo {author} {\bibfnamefont {D.}~\bibnamefont {Betto}}, \bibinfo {author} {\bibfnamefont {S.}~\bibnamefont {Ishida}}, \bibinfo {author} {\bibfnamefont {H.}~\bibnamefont {Eisaki}}, \bibinfo {author} {\bibfnamefont {C.}~\bibnamefont {Hu}}, \bibinfo {author} {\bibfnamefont {X.}~\bibnamefont {Zhou}}, \bibinfo {author} {\bibfnamefont {M.~W.}\ \bibnamefont {Haverkort}}, \bibinfo {author} {\bibfnamefont {N.~B.}\ \bibnamefont {Brookes}},
  \bibinfo {author} {\bibfnamefont {L.}~\bibnamefont {Braicovich}}, \bibinfo {author} {\bibfnamefont {G.}~\bibnamefont {Ghiringhelli}}, \bibinfo {author} {\bibfnamefont {M.}~\bibnamefont {Le~Tacon}},\ and\ \bibinfo {author} {\bibfnamefont {B.}~\bibnamefont {Keimer}},\ }\bibfield  {title} {\bibinfo {title} {Probing the energy gap of high-temperature cuprate superconductors by resonant inelastic x-ray scattering},\ }\href {https://doi.org/10.1038/s41535-018-0139-7} {\bibfield  {journal} {\bibinfo  {journal} {npj Quantum Materials}\ }\textbf {\bibinfo {volume} {3}},\ \bibinfo {pages} {65} (\bibinfo {year} {2018})}\BibitemShut {NoStop}%
\bibitem [{\citenamefont {Merzoni}\ \emph {et~al.}(2024)\citenamefont {Merzoni}, \citenamefont {Martinelli}, \citenamefont {Braicovich}, \citenamefont {Brookes}, \citenamefont {Lombardi}, \citenamefont {Rosa}, \citenamefont {Arpaia}, \citenamefont {Moretti~Sala},\ and\ \citenamefont {Ghiringhelli}}]{Merzoni2024}%
  \BibitemOpen
  \bibfield  {author} {\bibinfo {author} {\bibfnamefont {G.}~\bibnamefont {Merzoni}}, \bibinfo {author} {\bibfnamefont {L.}~\bibnamefont {Martinelli}}, \bibinfo {author} {\bibfnamefont {L.}~\bibnamefont {Braicovich}}, \bibinfo {author} {\bibfnamefont {N.~B.}\ \bibnamefont {Brookes}}, \bibinfo {author} {\bibfnamefont {F.}~\bibnamefont {Lombardi}}, \bibinfo {author} {\bibfnamefont {F.}~\bibnamefont {Rosa}}, \bibinfo {author} {\bibfnamefont {R.}~\bibnamefont {Arpaia}}, \bibinfo {author} {\bibfnamefont {M.}~\bibnamefont {Moretti~Sala}},\ and\ \bibinfo {author} {\bibfnamefont {G.}~\bibnamefont {Ghiringhelli}},\ }\bibfield  {title} {\bibinfo {title} {{Charge response function probed by resonant inelastic x-ray scattering: Signature of electronic gaps of ${\mathrm{YBa}}_{2}{\mathrm{Cu}}_{3}{\mathrm{O}}_{7\ensuremath{-}\ensuremath{\delta}}$}},\ }\href {https://doi.org/10.1103/PhysRevB.109.184506} {\bibfield  {journal} {\bibinfo  {journal} {Phys. Rev. B}\ }\textbf {\bibinfo {volume} {109}},\ \bibinfo {pages} {184506}
  (\bibinfo {year} {2024})}\BibitemShut {NoStop}%
\bibitem [{\citenamefont {Jia}\ \emph {et~al.}(2016)\citenamefont {Jia}, \citenamefont {Wohlfeld}, \citenamefont {Wang}, \citenamefont {Moritz},\ and\ \citenamefont {Devereaux}}]{Jia2016}%
  \BibitemOpen
  \bibfield  {author} {\bibinfo {author} {\bibfnamefont {C.}~\bibnamefont {Jia}}, \bibinfo {author} {\bibfnamefont {K.}~\bibnamefont {Wohlfeld}}, \bibinfo {author} {\bibfnamefont {Y.}~\bibnamefont {Wang}}, \bibinfo {author} {\bibfnamefont {B.}~\bibnamefont {Moritz}},\ and\ \bibinfo {author} {\bibfnamefont {T.~P.}\ \bibnamefont {Devereaux}},\ }\bibfield  {title} {\bibinfo {title} {Using rixs to uncover elementary charge and spin excitations},\ }\href {https://doi.org/10.1103/PhysRevX.6.021020} {\bibfield  {journal} {\bibinfo  {journal} {Phys. Rev. X}\ }\textbf {\bibinfo {volume} {6}},\ \bibinfo {pages} {021020} (\bibinfo {year} {2016})}\BibitemShut {NoStop}%
\bibitem [{sm()}]{sm}%
  \BibitemOpen
  \href@noop {} {}\bibinfo {note} {See Supplementary Material for details at ...}\BibitemShut {Stop}%
\bibitem [{\citenamefont {Norman}\ \emph {et~al.}(1995)\citenamefont {Norman}, \citenamefont {Randeria}, \citenamefont {Ding},\ and\ \citenamefont {Campuzano}}]{Norman1995}%
  \BibitemOpen
  \bibfield  {author} {\bibinfo {author} {\bibfnamefont {M.~R.}\ \bibnamefont {Norman}}, \bibinfo {author} {\bibfnamefont {M.}~\bibnamefont {Randeria}}, \bibinfo {author} {\bibfnamefont {H.}~\bibnamefont {Ding}},\ and\ \bibinfo {author} {\bibfnamefont {J.~C.}\ \bibnamefont {Campuzano}},\ }\bibfield  {title} {\bibinfo {title} {{Phenomenological models for the gap anisotropy of ${\mathrm{Bi}}_{2}$${\mathrm{Sr}}_{2}$${\mathrm{CaCu}}_{2}$${\mathrm{O}}_{8}$ as measured by angle-resolved photoemission spectroscopy}},\ }\href {https://doi.org/10.1103/PhysRevB.52.615} {\bibfield  {journal} {\bibinfo  {journal} {Phys. Rev. B}\ }\textbf {\bibinfo {volume} {52}},\ \bibinfo {pages} {615} (\bibinfo {year} {1995})}\BibitemShut {NoStop}%
\bibitem [{\citenamefont {Lee}\ \emph {et~al.}(2007)\citenamefont {Lee}, \citenamefont {Vishik}, \citenamefont {Tanaka}, \citenamefont {Lu}, \citenamefont {Sasagawa}, \citenamefont {Nagaosa}, \citenamefont {Devereaux}, \citenamefont {Hussain},\ and\ \citenamefont {Shen}}]{Lee2007}%
  \BibitemOpen
  \bibfield  {author} {\bibinfo {author} {\bibfnamefont {W.~S.}\ \bibnamefont {Lee}}, \bibinfo {author} {\bibfnamefont {I.~M.}\ \bibnamefont {Vishik}}, \bibinfo {author} {\bibfnamefont {K.}~\bibnamefont {Tanaka}}, \bibinfo {author} {\bibfnamefont {D.~H.}\ \bibnamefont {Lu}}, \bibinfo {author} {\bibfnamefont {T.}~\bibnamefont {Sasagawa}}, \bibinfo {author} {\bibfnamefont {N.}~\bibnamefont {Nagaosa}}, \bibinfo {author} {\bibfnamefont {T.~P.}\ \bibnamefont {Devereaux}}, \bibinfo {author} {\bibfnamefont {Z.}~\bibnamefont {Hussain}},\ and\ \bibinfo {author} {\bibfnamefont {Z.-X.}\ \bibnamefont {Shen}},\ }\bibfield  {title} {\bibinfo {title} {{Abrupt onset of a second energy gap at the superconducting transition of underdoped Bi2212}},\ }\href {https://doi.org/10.1038/nature06219} {\bibfield  {journal} {\bibinfo  {journal} {Nature}\ }\textbf {\bibinfo {volume} {450}},\ \bibinfo {pages} {81} (\bibinfo {year} {2007})}\BibitemShut {NoStop}%
\bibitem [{\citenamefont {Kondo}\ \emph {et~al.}(2009)\citenamefont {Kondo}, \citenamefont {Khasanov}, \citenamefont {Takeuchi}, \citenamefont {Schmalian},\ and\ \citenamefont {Kaminski}}]{Kondo2009}%
  \BibitemOpen
  \bibfield  {author} {\bibinfo {author} {\bibfnamefont {T.}~\bibnamefont {Kondo}}, \bibinfo {author} {\bibfnamefont {R.}~\bibnamefont {Khasanov}}, \bibinfo {author} {\bibfnamefont {T.}~\bibnamefont {Takeuchi}}, \bibinfo {author} {\bibfnamefont {J.}~\bibnamefont {Schmalian}},\ and\ \bibinfo {author} {\bibfnamefont {A.}~\bibnamefont {Kaminski}},\ }\bibfield  {title} {\bibinfo {title} {Competition between the pseudogap and superconductivity in the high-tc copper oxides},\ }\href {https://doi.org/10.1038/nature07644} {\bibfield  {journal} {\bibinfo  {journal} {Nature}\ }\textbf {\bibinfo {volume} {457}},\ \bibinfo {pages} {296} (\bibinfo {year} {2009})}\BibitemShut {NoStop}%
\bibitem [{\citenamefont {Valla}\ \emph {et~al.}(2020)\citenamefont {Valla}, \citenamefont {Drozdov},\ and\ \citenamefont {Gu}}]{Valla2020}%
  \BibitemOpen
  \bibfield  {author} {\bibinfo {author} {\bibfnamefont {T.}~\bibnamefont {Valla}}, \bibinfo {author} {\bibfnamefont {I.~K.}\ \bibnamefont {Drozdov}},\ and\ \bibinfo {author} {\bibfnamefont {G.~D.}\ \bibnamefont {Gu}},\ }\bibfield  {title} {\bibinfo {title} {{Disappearance of superconductivity due to vanishing coupling in the overdoped Bi$_2$Sr$_2$CaCu$_2$O$_{8+\delta}$}},\ }\href {https://doi.org/10.1038/s41467-020-14282-4} {\bibfield  {journal} {\bibinfo  {journal} {Nature Communications}\ }\textbf {\bibinfo {volume} {11}},\ \bibinfo {pages} {569} (\bibinfo {year} {2020})}\BibitemShut {NoStop}%
\bibitem [{\citenamefont {Vishik}\ \emph {et~al.}(2009)\citenamefont {Vishik}, \citenamefont {Nowadnick}, \citenamefont {Lee}, \citenamefont {Shen}, \citenamefont {Moritz}, \citenamefont {Devereaux}, \citenamefont {Tanaka}, \citenamefont {Sasagawa},\ and\ \citenamefont {Fujii}}]{Vishik2009}%
  \BibitemOpen
  \bibfield  {author} {\bibinfo {author} {\bibfnamefont {I.~M.}\ \bibnamefont {Vishik}}, \bibinfo {author} {\bibfnamefont {E.~A.}\ \bibnamefont {Nowadnick}}, \bibinfo {author} {\bibfnamefont {W.~S.}\ \bibnamefont {Lee}}, \bibinfo {author} {\bibfnamefont {Z.~X.}\ \bibnamefont {Shen}}, \bibinfo {author} {\bibfnamefont {B.}~\bibnamefont {Moritz}}, \bibinfo {author} {\bibfnamefont {T.~P.}\ \bibnamefont {Devereaux}}, \bibinfo {author} {\bibfnamefont {K.}~\bibnamefont {Tanaka}}, \bibinfo {author} {\bibfnamefont {T.}~\bibnamefont {Sasagawa}},\ and\ \bibinfo {author} {\bibfnamefont {T.}~\bibnamefont {Fujii}},\ }\bibfield  {title} {\bibinfo {title} {{A momentum-dependent perspective on quasiparticle interference in Bi$_2$Sr$_2$CaCu$_2$O$_8$}},\ }\href {https://doi.org/10.1038/nphys1375} {\bibfield  {journal} {\bibinfo  {journal} {Nature Physics}\ }\textbf {\bibinfo {volume} {5}},\ \bibinfo {pages} {718} (\bibinfo {year} {2009})}\BibitemShut {NoStop}%
\bibitem [{\citenamefont {Vishik}\ \emph {et~al.}(2012)\citenamefont {Vishik}, \citenamefont {Hashimoto}, \citenamefont {He}, \citenamefont {Lee}, \citenamefont {Schmitt}, \citenamefont {Lu}, \citenamefont {Moore}, \citenamefont {Zhang}, \citenamefont {Meevasana}, \citenamefont {Sasagawa}, \citenamefont {Uchida}, \citenamefont {Fujita}, \citenamefont {Ishida}, \citenamefont {Ishikado}, \citenamefont {Yoshida}, \citenamefont {Eisaki}, \citenamefont {Hussain}, \citenamefont {Devereaux},\ and\ \citenamefont {Shen}}]{Vishik2012}%
  \BibitemOpen
  \bibfield  {author} {\bibinfo {author} {\bibfnamefont {I.~M.}\ \bibnamefont {Vishik}}, \bibinfo {author} {\bibfnamefont {M.}~\bibnamefont {Hashimoto}}, \bibinfo {author} {\bibfnamefont {R.-H.}\ \bibnamefont {He}}, \bibinfo {author} {\bibfnamefont {W.-S.}\ \bibnamefont {Lee}}, \bibinfo {author} {\bibfnamefont {F.}~\bibnamefont {Schmitt}}, \bibinfo {author} {\bibfnamefont {D.}~\bibnamefont {Lu}}, \bibinfo {author} {\bibfnamefont {R.~G.}\ \bibnamefont {Moore}}, \bibinfo {author} {\bibfnamefont {C.}~\bibnamefont {Zhang}}, \bibinfo {author} {\bibfnamefont {W.}~\bibnamefont {Meevasana}}, \bibinfo {author} {\bibfnamefont {T.}~\bibnamefont {Sasagawa}}, \bibinfo {author} {\bibfnamefont {S.}~\bibnamefont {Uchida}}, \bibinfo {author} {\bibfnamefont {K.}~\bibnamefont {Fujita}}, \bibinfo {author} {\bibfnamefont {S.}~\bibnamefont {Ishida}}, \bibinfo {author} {\bibfnamefont {M.}~\bibnamefont {Ishikado}}, \bibinfo {author} {\bibfnamefont {Y.}~\bibnamefont {Yoshida}}, \bibinfo {author} {\bibfnamefont {H.}~\bibnamefont
  {Eisaki}}, \bibinfo {author} {\bibfnamefont {Z.}~\bibnamefont {Hussain}}, \bibinfo {author} {\bibfnamefont {T.~P.}\ \bibnamefont {Devereaux}},\ and\ \bibinfo {author} {\bibfnamefont {Z.-X.}\ \bibnamefont {Shen}},\ }\bibfield  {title} {\bibinfo {title} {Phase competition in trisected superconducting dome},\ }\href {https://doi.org/10.1073/pnas.1209471109} {\bibfield  {journal} {\bibinfo  {journal} {Proceedings of the National Academy of Sciences}\ }\textbf {\bibinfo {volume} {109}},\ \bibinfo {pages} {18332} (\bibinfo {year} {2012})}\BibitemShut {NoStop}%
\bibitem [{\citenamefont {Onari}\ \emph {et~al.}(2010)\citenamefont {Onari}, \citenamefont {Kontani},\ and\ \citenamefont {Sato}}]{Onari2010}%
  \BibitemOpen
  \bibfield  {author} {\bibinfo {author} {\bibfnamefont {S.}~\bibnamefont {Onari}}, \bibinfo {author} {\bibfnamefont {H.}~\bibnamefont {Kontani}},\ and\ \bibinfo {author} {\bibfnamefont {M.}~\bibnamefont {Sato}},\ }\bibfield  {title} {\bibinfo {title} {{Structure of neutron-scattering peaks in both ${s}_{++}$-wave and ${s}_{\ifmmode\pm\else\textpm\fi{}}$-wave states of an iron pnictide superconductor}},\ }\href {https://doi.org/10.1103/PhysRevB.81.060504} {\bibfield  {journal} {\bibinfo  {journal} {Phys. Rev. B}\ }\textbf {\bibinfo {volume} {81}},\ \bibinfo {pages} {060504} (\bibinfo {year} {2010})}\BibitemShut {NoStop}%
\bibitem [{\citenamefont {Hashimoto}\ \emph {et~al.}(2014)\citenamefont {Hashimoto}, \citenamefont {Vishik}, \citenamefont {He}, \citenamefont {Devereaux},\ and\ \citenamefont {Shen}}]{Hashimoto2014}%
  \BibitemOpen
  \bibfield  {author} {\bibinfo {author} {\bibfnamefont {M.}~\bibnamefont {Hashimoto}}, \bibinfo {author} {\bibfnamefont {I.~M.}\ \bibnamefont {Vishik}}, \bibinfo {author} {\bibfnamefont {R.-H.}\ \bibnamefont {He}}, \bibinfo {author} {\bibfnamefont {T.~P.}\ \bibnamefont {Devereaux}},\ and\ \bibinfo {author} {\bibfnamefont {Z.-X.}\ \bibnamefont {Shen}},\ }\bibfield  {title} {\bibinfo {title} {Energy gaps in high-transition-temperature cuprate superconductors},\ }\href {https://doi.org/10.1038/nphys3009} {\bibfield  {journal} {\bibinfo  {journal} {Nature Physics}\ }\textbf {\bibinfo {volume} {10}},\ \bibinfo {pages} {483} (\bibinfo {year} {2014})}\BibitemShut {NoStop}%
\bibitem [{\citenamefont {Dvorak}\ \emph {et~al.}(2016)\citenamefont {Dvorak}, \citenamefont {Jarrige}, \citenamefont {Bisogni}, \citenamefont {Coburn},\ and\ \citenamefont {Leonhardt}}]{Dvorak16}%
  \BibitemOpen
  \bibfield  {author} {\bibinfo {author} {\bibfnamefont {J.}~\bibnamefont {Dvorak}}, \bibinfo {author} {\bibfnamefont {I.}~\bibnamefont {Jarrige}}, \bibinfo {author} {\bibfnamefont {V.}~\bibnamefont {Bisogni}}, \bibinfo {author} {\bibfnamefont {S.}~\bibnamefont {Coburn}},\ and\ \bibinfo {author} {\bibfnamefont {W.}~\bibnamefont {Leonhardt}},\ }\bibfield  {title} {\bibinfo {title} {Towards 10 {meV} resolution: {The} design of an ultrahigh resolution soft {X}-ray {RIXS} spectrometer},\ }\href {https://doi.org/10.1063/1.4964847} {\bibfield  {journal} {\bibinfo  {journal} {Review of Scientific Instruments}\ }\textbf {\bibinfo {volume} {87}},\ \bibinfo {pages} {115109} (\bibinfo {year} {2016})}\BibitemShut {NoStop}%
\bibitem [{\citenamefont {Dean}\ \emph {et~al.}(2013)\citenamefont {Dean}, \citenamefont {James}, \citenamefont {Springell}, \citenamefont {Liu}, \citenamefont {Monney}, \citenamefont {Zhou}, \citenamefont {Konik}, \citenamefont {Wen}, \citenamefont {Xu}, \citenamefont {Gu}, \citenamefont {Strocov}, \citenamefont {Schmitt},\ and\ \citenamefont {Hill}}]{Dean2013}%
  \BibitemOpen
  \bibfield  {author} {\bibinfo {author} {\bibfnamefont {M.~P.~M.}\ \bibnamefont {Dean}}, \bibinfo {author} {\bibfnamefont {A.~J.~A.}\ \bibnamefont {James}}, \bibinfo {author} {\bibfnamefont {R.~S.}\ \bibnamefont {Springell}}, \bibinfo {author} {\bibfnamefont {X.}~\bibnamefont {Liu}}, \bibinfo {author} {\bibfnamefont {C.}~\bibnamefont {Monney}}, \bibinfo {author} {\bibfnamefont {K.~J.}\ \bibnamefont {Zhou}}, \bibinfo {author} {\bibfnamefont {R.~M.}\ \bibnamefont {Konik}}, \bibinfo {author} {\bibfnamefont {J.~S.}\ \bibnamefont {Wen}}, \bibinfo {author} {\bibfnamefont {Z.~J.}\ \bibnamefont {Xu}}, \bibinfo {author} {\bibfnamefont {G.~D.}\ \bibnamefont {Gu}}, \bibinfo {author} {\bibfnamefont {V.~N.}\ \bibnamefont {Strocov}}, \bibinfo {author} {\bibfnamefont {T.}~\bibnamefont {Schmitt}},\ and\ \bibinfo {author} {\bibfnamefont {J.~P.}\ \bibnamefont {Hill}},\ }\bibfield  {title} {\bibinfo {title} {{High-Energy Magnetic Excitations in the Cuprate Superconductor
  ${\mathrm{Bi}}_{2}{\mathrm{Sr}}_{2}{\mathrm{CaCu}}_{2}{\mathbf{O}}_{8\mathbf{+}\ensuremath{\delta}}$: Towards a Unified Description of Its Electronic and Magnetic Degrees of Freedom}},\ }\href {https://doi.org/10.1103/PhysRevLett.110.147001} {\bibfield  {journal} {\bibinfo  {journal} {Phys. Rev. Lett.}\ }\textbf {\bibinfo {volume} {110}},\ \bibinfo {pages} {147001} (\bibinfo {year} {2013})}\BibitemShut {NoStop}%
\bibitem [{\citenamefont {Dean}\ \emph {et~al.}(2014)\citenamefont {Dean}, \citenamefont {James}, \citenamefont {Walters}, \citenamefont {Bisogni}, \citenamefont {Jarrige}, \citenamefont {H\"ucker}, \citenamefont {Giannini}, \citenamefont {Fujita}, \citenamefont {Pelliciari}, \citenamefont {Huang}, \citenamefont {Konik}, \citenamefont {Schmitt},\ and\ \citenamefont {Hill}}]{Dean2014}%
  \BibitemOpen
  \bibfield  {author} {\bibinfo {author} {\bibfnamefont {M.~P.~M.}\ \bibnamefont {Dean}}, \bibinfo {author} {\bibfnamefont {A.~J.~A.}\ \bibnamefont {James}}, \bibinfo {author} {\bibfnamefont {A.~C.}\ \bibnamefont {Walters}}, \bibinfo {author} {\bibfnamefont {V.}~\bibnamefont {Bisogni}}, \bibinfo {author} {\bibfnamefont {I.}~\bibnamefont {Jarrige}}, \bibinfo {author} {\bibfnamefont {M.}~\bibnamefont {H\"ucker}}, \bibinfo {author} {\bibfnamefont {E.}~\bibnamefont {Giannini}}, \bibinfo {author} {\bibfnamefont {M.}~\bibnamefont {Fujita}}, \bibinfo {author} {\bibfnamefont {J.}~\bibnamefont {Pelliciari}}, \bibinfo {author} {\bibfnamefont {Y.~B.}\ \bibnamefont {Huang}}, \bibinfo {author} {\bibfnamefont {R.~M.}\ \bibnamefont {Konik}}, \bibinfo {author} {\bibfnamefont {T.}~\bibnamefont {Schmitt}},\ and\ \bibinfo {author} {\bibfnamefont {J.~P.}\ \bibnamefont {Hill}},\ }\bibfield  {title} {\bibinfo {title} {Itinerant effects and enhanced magnetic interactions in bi-based multilayer cuprates},\ }\href
  {https://doi.org/10.1103/PhysRevB.90.220506} {\bibfield  {journal} {\bibinfo  {journal} {Phys. Rev. B}\ }\textbf {\bibinfo {volume} {90}},\ \bibinfo {pages} {220506} (\bibinfo {year} {2014})}\BibitemShut {NoStop}%
\bibitem [{\citenamefont {Dean}(2015)}]{Dean2015}%
  \BibitemOpen
  \bibfield  {author} {\bibinfo {author} {\bibfnamefont {M.}~\bibnamefont {Dean}},\ }\bibfield  {title} {\bibinfo {title} {Insights into the high temperature superconducting cuprates from resonant inelastic x-ray scattering},\ }\href {https://doi.org/https://doi.org/10.1016/j.jmmm.2014.03.057} {\bibfield  {journal} {\bibinfo  {journal} {Journal of Magnetism and Magnetic Materials}\ }\textbf {\bibinfo {volume} {376}},\ \bibinfo {pages} {3} (\bibinfo {year} {2015})}\BibitemShut {NoStop}%
\bibitem [{\citenamefont {Chaix}\ \emph {et~al.}(2017)\citenamefont {Chaix}, \citenamefont {Ghiringhelli}, \citenamefont {Peng}, \citenamefont {Hashimoto}, \citenamefont {Moritz}, \citenamefont {Kummer}, \citenamefont {Brookes}, \citenamefont {He}, \citenamefont {Chen}, \citenamefont {Ishida}, \citenamefont {Yoshida}, \citenamefont {Eisaki}, \citenamefont {Salluzzo}, \citenamefont {Braicovich}, \citenamefont {Shen}, \citenamefont {Devereaux},\ and\ \citenamefont {Lee}}]{Chaix2017}%
  \BibitemOpen
  \bibfield  {author} {\bibinfo {author} {\bibfnamefont {L.}~\bibnamefont {Chaix}}, \bibinfo {author} {\bibfnamefont {G.}~\bibnamefont {Ghiringhelli}}, \bibinfo {author} {\bibfnamefont {Y.~Y.}\ \bibnamefont {Peng}}, \bibinfo {author} {\bibfnamefont {M.}~\bibnamefont {Hashimoto}}, \bibinfo {author} {\bibfnamefont {B.}~\bibnamefont {Moritz}}, \bibinfo {author} {\bibfnamefont {K.}~\bibnamefont {Kummer}}, \bibinfo {author} {\bibfnamefont {N.~B.}\ \bibnamefont {Brookes}}, \bibinfo {author} {\bibfnamefont {Y.}~\bibnamefont {He}}, \bibinfo {author} {\bibfnamefont {S.}~\bibnamefont {Chen}}, \bibinfo {author} {\bibfnamefont {S.}~\bibnamefont {Ishida}}, \bibinfo {author} {\bibfnamefont {Y.}~\bibnamefont {Yoshida}}, \bibinfo {author} {\bibfnamefont {H.}~\bibnamefont {Eisaki}}, \bibinfo {author} {\bibfnamefont {M.}~\bibnamefont {Salluzzo}}, \bibinfo {author} {\bibfnamefont {L.}~\bibnamefont {Braicovich}}, \bibinfo {author} {\bibfnamefont {Z.-X.}\ \bibnamefont {Shen}}, \bibinfo {author} {\bibfnamefont {T.~P.}\
  \bibnamefont {Devereaux}},\ and\ \bibinfo {author} {\bibfnamefont {W.-S.}\ \bibnamefont {Lee}},\ }\bibfield  {title} {\bibinfo {title} {{Dispersive charge density wave excitations in ${\mathrm{Bi}}_{2}{\mathrm{Sr}}_{2}{\mathrm{CaCu}}_{2}{\mathbf{O}}_{8\mathbf{+}\ensuremath{\delta}}$: Towards a Unified Description of Its Electronic and Magnetic Degrees of Freedom}},\ }\href {https://doi.org/10.1038/nphys4157} {\bibfield  {journal} {\bibinfo  {journal} {Nature Physics}\ }\textbf {\bibinfo {volume} {13}},\ \bibinfo {pages} {952} (\bibinfo {year} {2017})}\BibitemShut {NoStop}%
\bibitem [{\citenamefont {Li}\ \emph {et~al.}(2020)\citenamefont {Li}, \citenamefont {Nag}, \citenamefont {Pelliciari}, \citenamefont {Robarts}, \citenamefont {Walters}, \citenamefont {Garcia-Fernandez}, \citenamefont {Eisaki}, \citenamefont {Song}, \citenamefont {Ding}, \citenamefont {Johnston}, \citenamefont {Comin},\ and\ \citenamefont {Zhou}}]{Jiemin2020}%
  \BibitemOpen
  \bibfield  {author} {\bibinfo {author} {\bibfnamefont {J.}~\bibnamefont {Li}}, \bibinfo {author} {\bibfnamefont {A.}~\bibnamefont {Nag}}, \bibinfo {author} {\bibfnamefont {J.}~\bibnamefont {Pelliciari}}, \bibinfo {author} {\bibfnamefont {H.}~\bibnamefont {Robarts}}, \bibinfo {author} {\bibfnamefont {A.}~\bibnamefont {Walters}}, \bibinfo {author} {\bibfnamefont {M.}~\bibnamefont {Garcia-Fernandez}}, \bibinfo {author} {\bibfnamefont {H.}~\bibnamefont {Eisaki}}, \bibinfo {author} {\bibfnamefont {D.}~\bibnamefont {Song}}, \bibinfo {author} {\bibfnamefont {H.}~\bibnamefont {Ding}}, \bibinfo {author} {\bibfnamefont {S.}~\bibnamefont {Johnston}}, \bibinfo {author} {\bibfnamefont {R.}~\bibnamefont {Comin}},\ and\ \bibinfo {author} {\bibfnamefont {K.-J.}\ \bibnamefont {Zhou}},\ }\bibfield  {title} {\bibinfo {title} {{Multiorbital charge-density wave excitations and concomitant phonon anomalies in Bi$_2$Sr$_2$LaCuO$_{6+\delta}$}},\ }\href {https://doi.org/10.1073/pnas.2001755117} {\bibfield  {journal} {\bibinfo
  {journal} {Proceedings of the National Academy of Sciences}\ }\textbf {\bibinfo {volume} {117}},\ \bibinfo {pages} {16219} (\bibinfo {year} {2020})}\BibitemShut {NoStop}%
\bibitem [{\citenamefont {Devereaux}\ \emph {et~al.}(2016)\citenamefont {Devereaux}, \citenamefont {Shvaika}, \citenamefont {Wu}, \citenamefont {Wohlfeld}, \citenamefont {Jia}, \citenamefont {Wang}, \citenamefont {Moritz}, \citenamefont {Chaix}, \citenamefont {Lee}, \citenamefont {Shen}, \citenamefont {Ghiringhelli},\ and\ \citenamefont {Braicovich}}]{Devereaux2016}%
  \BibitemOpen
  \bibfield  {author} {\bibinfo {author} {\bibfnamefont {T.~P.}\ \bibnamefont {Devereaux}}, \bibinfo {author} {\bibfnamefont {A.~M.}\ \bibnamefont {Shvaika}}, \bibinfo {author} {\bibfnamefont {K.}~\bibnamefont {Wu}}, \bibinfo {author} {\bibfnamefont {K.}~\bibnamefont {Wohlfeld}}, \bibinfo {author} {\bibfnamefont {C.~J.}\ \bibnamefont {Jia}}, \bibinfo {author} {\bibfnamefont {Y.}~\bibnamefont {Wang}}, \bibinfo {author} {\bibfnamefont {B.}~\bibnamefont {Moritz}}, \bibinfo {author} {\bibfnamefont {L.}~\bibnamefont {Chaix}}, \bibinfo {author} {\bibfnamefont {W.-S.}\ \bibnamefont {Lee}}, \bibinfo {author} {\bibfnamefont {Z.-X.}\ \bibnamefont {Shen}}, \bibinfo {author} {\bibfnamefont {G.}~\bibnamefont {Ghiringhelli}},\ and\ \bibinfo {author} {\bibfnamefont {L.}~\bibnamefont {Braicovich}},\ }\bibfield  {title} {\bibinfo {title} {Directly characterizing the relative strength and momentum dependence of electron-phonon coupling using resonant inelastic x-ray scattering},\ }\href
  {https://doi.org/10.1103/PhysRevX.6.041019} {\bibfield  {journal} {\bibinfo  {journal} {Phys. Rev. X}\ }\textbf {\bibinfo {volume} {6}},\ \bibinfo {pages} {041019} (\bibinfo {year} {2016})}\BibitemShut {NoStop}%
\bibitem [{\citenamefont {Robarts}\ \emph {et~al.}(2019)\citenamefont {Robarts}, \citenamefont {Barth\'elemy}, \citenamefont {Kummer}, \citenamefont {Garc\'{\i}a-Fern\'andez}, \citenamefont {Li}, \citenamefont {Nag}, \citenamefont {Walters}, \citenamefont {Zhou},\ and\ \citenamefont {Hayden}}]{Robarts2019}%
  \BibitemOpen
  \bibfield  {author} {\bibinfo {author} {\bibfnamefont {H.~C.}\ \bibnamefont {Robarts}}, \bibinfo {author} {\bibfnamefont {M.}~\bibnamefont {Barth\'elemy}}, \bibinfo {author} {\bibfnamefont {K.}~\bibnamefont {Kummer}}, \bibinfo {author} {\bibfnamefont {M.}~\bibnamefont {Garc\'{\i}a-Fern\'andez}}, \bibinfo {author} {\bibfnamefont {J.}~\bibnamefont {Li}}, \bibinfo {author} {\bibfnamefont {A.}~\bibnamefont {Nag}}, \bibinfo {author} {\bibfnamefont {A.~C.}\ \bibnamefont {Walters}}, \bibinfo {author} {\bibfnamefont {K.~J.}\ \bibnamefont {Zhou}},\ and\ \bibinfo {author} {\bibfnamefont {S.~M.}\ \bibnamefont {Hayden}},\ }\bibfield  {title} {\bibinfo {title} {{Anisotropic damping and wave vector dependent susceptibility of the spin fluctuations in ${\mathrm{La}}_{2\ensuremath{-}x}{\mathrm{Sr}}_{x}{\mathrm{CuO}}_{4}$ studied by resonant inelastic x-ray scattering}},\ }\href {https://doi.org/10.1103/PhysRevB.100.214510} {\bibfield  {journal} {\bibinfo  {journal} {Phys. Rev. B}\ }\textbf {\bibinfo {volume} {100}},\
  \bibinfo {pages} {214510} (\bibinfo {year} {2019})}\BibitemShut {NoStop}%
\bibitem [{\citenamefont {Hepting}\ \emph {et~al.}(2018)\citenamefont {Hepting}, \citenamefont {Chaix}, \citenamefont {Huang}, \citenamefont {Fumagalli}, \citenamefont {Peng}, \citenamefont {Moritz}, \citenamefont {Kummer}, \citenamefont {Brookes}, \citenamefont {Lee}, \citenamefont {Hashimoto}, \citenamefont {Sarkar}, \citenamefont {He}, \citenamefont {Rotundu}, \citenamefont {Lee}, \citenamefont {Greene}, \citenamefont {Braicovich}, \citenamefont {Ghiringhelli}, \citenamefont {Shen}, \citenamefont {evereaux},\ and\ \citenamefont {Lee}}]{Hepting2018}%
  \BibitemOpen
  \bibfield  {author} {\bibinfo {author} {\bibfnamefont {M.}~\bibnamefont {Hepting}}, \bibinfo {author} {\bibfnamefont {L.}~\bibnamefont {Chaix}}, \bibinfo {author} {\bibfnamefont {E.~W.}\ \bibnamefont {Huang}}, \bibinfo {author} {\bibfnamefont {R.}~\bibnamefont {Fumagalli}}, \bibinfo {author} {\bibfnamefont {Y.~Y.}\ \bibnamefont {Peng}}, \bibinfo {author} {\bibfnamefont {B.}~\bibnamefont {Moritz}}, \bibinfo {author} {\bibfnamefont {K.}~\bibnamefont {Kummer}}, \bibinfo {author} {\bibfnamefont {N.~B.}\ \bibnamefont {Brookes}}, \bibinfo {author} {\bibfnamefont {W.~C.}\ \bibnamefont {Lee}}, \bibinfo {author} {\bibfnamefont {M.}~\bibnamefont {Hashimoto}}, \bibinfo {author} {\bibfnamefont {T.}~\bibnamefont {Sarkar}}, \bibinfo {author} {\bibfnamefont {J.-F.}\ \bibnamefont {He}}, \bibinfo {author} {\bibfnamefont {C.~R.}\ \bibnamefont {Rotundu}}, \bibinfo {author} {\bibfnamefont {Y.~S.}\ \bibnamefont {Lee}}, \bibinfo {author} {\bibfnamefont {R.~L.}\ \bibnamefont {Greene}}, \bibinfo {author} {\bibfnamefont
  {L.}~\bibnamefont {Braicovich}}, \bibinfo {author} {\bibfnamefont {G.}~\bibnamefont {Ghiringhelli}}, \bibinfo {author} {\bibfnamefont {Z.~X.}\ \bibnamefont {Shen}}, \bibinfo {author} {\bibfnamefont {T.~P.}\ \bibnamefont {evereaux}},\ and\ \bibinfo {author} {\bibfnamefont {W.~S.}\ \bibnamefont {Lee}},\ }\bibfield  {title} {\bibinfo {title} {{Three-dimensional collective charge excitations in electron-doped copper oxide superconductors}},\ }\href {https://doi.org/10.1038/s41586-018-0648-3} {\bibfield  {journal} {\bibinfo  {journal} {Nature}\ }\textbf {\bibinfo {volume} {563}},\ \bibinfo {pages} {374} (\bibinfo {year} {2018})}\BibitemShut {NoStop}%
\bibitem [{\citenamefont {Nag}\ \emph {et~al.}(2020)\citenamefont {Nag}, \citenamefont {Zhu}, \citenamefont {Bejas}, \citenamefont {Li}, \citenamefont {Robarts}, \citenamefont {Yamase}, \citenamefont {Petsch}, \citenamefont {Song}, \citenamefont {Eisaki}, \citenamefont {Walters}, \citenamefont {Garc\'{\i}a-Fern\'andez}, \citenamefont {Greco}, \citenamefont {Hayden},\ and\ \citenamefont {Zhou}}]{Nag2020}%
  \BibitemOpen
  \bibfield  {author} {\bibinfo {author} {\bibfnamefont {A.}~\bibnamefont {Nag}}, \bibinfo {author} {\bibfnamefont {M.}~\bibnamefont {Zhu}}, \bibinfo {author} {\bibfnamefont {M.}~\bibnamefont {Bejas}}, \bibinfo {author} {\bibfnamefont {J.}~\bibnamefont {Li}}, \bibinfo {author} {\bibfnamefont {H.~C.}\ \bibnamefont {Robarts}}, \bibinfo {author} {\bibfnamefont {H.}~\bibnamefont {Yamase}}, \bibinfo {author} {\bibfnamefont {A.~N.}\ \bibnamefont {Petsch}}, \bibinfo {author} {\bibfnamefont {D.}~\bibnamefont {Song}}, \bibinfo {author} {\bibfnamefont {H.}~\bibnamefont {Eisaki}}, \bibinfo {author} {\bibfnamefont {A.~C.}\ \bibnamefont {Walters}}, \bibinfo {author} {\bibfnamefont {M.}~\bibnamefont {Garc\'{\i}a-Fern\'andez}}, \bibinfo {author} {\bibfnamefont {A.}~\bibnamefont {Greco}}, \bibinfo {author} {\bibfnamefont {S.~M.}\ \bibnamefont {Hayden}},\ and\ \bibinfo {author} {\bibfnamefont {K.-J.}\ \bibnamefont {Zhou}},\ }\bibfield  {title} {\bibinfo {title} {Detection of acoustic plasmons in hole-doped lanthanum and
  bismuth cuprate superconductors using resonant inelastic x-ray scattering},\ }\href {https://doi.org/10.1103/PhysRevLett.125.257002} {\bibfield  {journal} {\bibinfo  {journal} {Phys. Rev. Lett.}\ }\textbf {\bibinfo {volume} {125}},\ \bibinfo {pages} {257002} (\bibinfo {year} {2020})}\BibitemShut {NoStop}%
\bibitem [{\citenamefont {Nag}\ \emph {et~al.}(2024)\citenamefont {Nag}, \citenamefont {Zinni}, \citenamefont {Choi}, \citenamefont {Li}, \citenamefont {Tu}, \citenamefont {Walters}, \citenamefont {Agrestini}, \citenamefont {Hayden}, \citenamefont {Bejas}, \citenamefont {Lin}, \citenamefont {Yamase}, \citenamefont {Jin}, \citenamefont {Garc\'{\i}a-Fern\'andez}, \citenamefont {Fink}, \citenamefont {Greco},\ and\ \citenamefont {Zhou}}]{Nag2024}%
  \BibitemOpen
  \bibfield  {author} {\bibinfo {author} {\bibfnamefont {A.}~\bibnamefont {Nag}}, \bibinfo {author} {\bibfnamefont {L.}~\bibnamefont {Zinni}}, \bibinfo {author} {\bibfnamefont {J.}~\bibnamefont {Choi}}, \bibinfo {author} {\bibfnamefont {J.}~\bibnamefont {Li}}, \bibinfo {author} {\bibfnamefont {S.}~\bibnamefont {Tu}}, \bibinfo {author} {\bibfnamefont {A.~C.}\ \bibnamefont {Walters}}, \bibinfo {author} {\bibfnamefont {S.}~\bibnamefont {Agrestini}}, \bibinfo {author} {\bibfnamefont {S.~M.}\ \bibnamefont {Hayden}}, \bibinfo {author} {\bibfnamefont {M.}~\bibnamefont {Bejas}}, \bibinfo {author} {\bibfnamefont {Z.}~\bibnamefont {Lin}}, \bibinfo {author} {\bibfnamefont {H.}~\bibnamefont {Yamase}}, \bibinfo {author} {\bibfnamefont {K.}~\bibnamefont {Jin}}, \bibinfo {author} {\bibfnamefont {M.}~\bibnamefont {Garc\'{\i}a-Fern\'andez}}, \bibinfo {author} {\bibfnamefont {J.}~\bibnamefont {Fink}}, \bibinfo {author} {\bibfnamefont {A.}~\bibnamefont {Greco}},\ and\ \bibinfo {author} {\bibfnamefont {K.-J.}\ \bibnamefont
  {Zhou}},\ }\bibfield  {title} {\bibinfo {title} {{Impact of electron correlations on two-particle charge response in electron- and hole-doped cuprates}},\ }\href {https://doi.org/10.1103/PhysRevResearch.6.043184} {\bibfield  {journal} {\bibinfo  {journal} {Phys. Rev. Res.}\ }\textbf {\bibinfo {volume} {6}},\ \bibinfo {pages} {043184} (\bibinfo {year} {2024})}\BibitemShut {NoStop}%
\bibitem [{\citenamefont {Scott}\ \emph {et~al.}(2023)\citenamefont {Scott}, \citenamefont {Kisiel}, \citenamefont {Boyle}, \citenamefont {Basak}, \citenamefont {Jargot}, \citenamefont {Das}, \citenamefont {Agrestini}, \citenamefont {Garcia-Fernandez}, \citenamefont {Choi}, \citenamefont {Pelliciari}, \citenamefont {Li}, \citenamefont {Chuang}, \citenamefont {Zhong}, \citenamefont {Schneeloch}, \citenamefont {Gu}, \citenamefont {Légaré}, \citenamefont {Kemper}, \citenamefont {Zhou}, \citenamefont {Bisogni}, \citenamefont {Blanco-Canosa}, \citenamefont {Frano}, \citenamefont {Boschini},\ and\ \citenamefont {da~Silva~Neto}}]{Kirsty2023}%
  \BibitemOpen
  \bibfield  {author} {\bibinfo {author} {\bibfnamefont {K.}~\bibnamefont {Scott}}, \bibinfo {author} {\bibfnamefont {E.}~\bibnamefont {Kisiel}}, \bibinfo {author} {\bibfnamefont {T.~J.}\ \bibnamefont {Boyle}}, \bibinfo {author} {\bibfnamefont {R.}~\bibnamefont {Basak}}, \bibinfo {author} {\bibfnamefont {G.}~\bibnamefont {Jargot}}, \bibinfo {author} {\bibfnamefont {S.}~\bibnamefont {Das}}, \bibinfo {author} {\bibfnamefont {S.}~\bibnamefont {Agrestini}}, \bibinfo {author} {\bibfnamefont {M.}~\bibnamefont {Garcia-Fernandez}}, \bibinfo {author} {\bibfnamefont {J.}~\bibnamefont {Choi}}, \bibinfo {author} {\bibfnamefont {J.}~\bibnamefont {Pelliciari}}, \bibinfo {author} {\bibfnamefont {J.}~\bibnamefont {Li}}, \bibinfo {author} {\bibfnamefont {Y.-D.}\ \bibnamefont {Chuang}}, \bibinfo {author} {\bibfnamefont {R.}~\bibnamefont {Zhong}}, \bibinfo {author} {\bibfnamefont {J.~A.}\ \bibnamefont {Schneeloch}}, \bibinfo {author} {\bibfnamefont {G.}~\bibnamefont {Gu}}, \bibinfo {author} {\bibfnamefont {F.}~\bibnamefont
  {Légaré}}, \bibinfo {author} {\bibfnamefont {A.~F.}\ \bibnamefont {Kemper}}, \bibinfo {author} {\bibfnamefont {K.-J.}\ \bibnamefont {Zhou}}, \bibinfo {author} {\bibfnamefont {V.}~\bibnamefont {Bisogni}}, \bibinfo {author} {\bibfnamefont {S.}~\bibnamefont {Blanco-Canosa}}, \bibinfo {author} {\bibfnamefont {A.}~\bibnamefont {Frano}}, \bibinfo {author} {\bibfnamefont {F.}~\bibnamefont {Boschini}},\ and\ \bibinfo {author} {\bibfnamefont {E.~H.}\ \bibnamefont {da~Silva~Neto}},\ }\bibfield  {title} {\bibinfo {title} {{Low-energy quasi-circular electron correlations with charge order wavelength in Ba$_2$Sr$_2$CaCu$_2$O$_{8+\delta}$}},\ }\href {https://doi.org/10.1126/sciadv.adg3710} {\bibfield  {journal} {\bibinfo  {journal} {Science Advances}\ }\textbf {\bibinfo {volume} {9}},\ \bibinfo {pages} {eadg3710} (\bibinfo {year} {2023})}\BibitemShut {NoStop}%
\bibitem [{\citenamefont {Greco}\ \emph {et~al.}(2016)\citenamefont {Greco}, \citenamefont {Yamase},\ and\ \citenamefont {Bejas}}]{Greco2016}%
  \BibitemOpen
  \bibfield  {author} {\bibinfo {author} {\bibfnamefont {A.}~\bibnamefont {Greco}}, \bibinfo {author} {\bibfnamefont {H.}~\bibnamefont {Yamase}},\ and\ \bibinfo {author} {\bibfnamefont {M.}~\bibnamefont {Bejas}},\ }\bibfield  {title} {\bibinfo {title} {{Plasmon excitations in layered high-${T}_{c}$ cuprates}},\ }\href {https://doi.org/10.1103/PhysRevB.94.075139} {\bibfield  {journal} {\bibinfo  {journal} {Phys. Rev. B}\ }\textbf {\bibinfo {volume} {94}},\ \bibinfo {pages} {075139} (\bibinfo {year} {2016})}\BibitemShut {NoStop}%
\bibitem [{\citenamefont {Greco}\ \emph {et~al.}(2019)\citenamefont {Greco}, \citenamefont {Yamase},\ and\ \citenamefont {Bejas}}]{Greco2019}%
  \BibitemOpen
  \bibfield  {author} {\bibinfo {author} {\bibfnamefont {A.}~\bibnamefont {Greco}}, \bibinfo {author} {\bibfnamefont {H.}~\bibnamefont {Yamase}},\ and\ \bibinfo {author} {\bibfnamefont {M.}~\bibnamefont {Bejas}},\ }\bibfield  {title} {\bibinfo {title} {{Origin of high-energy charge excitations observed by resonant inelastic X-ray scattering in cuprate superconductors}},\ }\href {https://doi.org/10.1038/s42005-018-0099-z} {\bibfield  {journal} {\bibinfo  {journal} {Communications Physics}\ }\textbf {\bibinfo {volume} {2}},\ \bibinfo {pages} {3} (\bibinfo {year} {2019})}\BibitemShut {NoStop}%
\bibitem [{\citenamefont {Zhu}\ \emph {et~al.}(2021)\citenamefont {Zhu}, \citenamefont {Liao}, \citenamefont {Zhang}, \citenamefont {Xie}, \citenamefont {Meng}, \citenamefont {Liu}, \citenamefont {Bai}, \citenamefont {Ji}, \citenamefont {Zhang}, \citenamefont {Jiang}, \citenamefont {Zhong}, \citenamefont {Schneeloch}, \citenamefont {Gu}, \citenamefont {Gu}, \citenamefont {Ma}, \citenamefont {Zhang},\ and\ \citenamefont {Xue}}]{Zhu2021}%
  \BibitemOpen
  \bibfield  {author} {\bibinfo {author} {\bibfnamefont {Y.}~\bibnamefont {Zhu}}, \bibinfo {author} {\bibfnamefont {M.}~\bibnamefont {Liao}}, \bibinfo {author} {\bibfnamefont {Q.}~\bibnamefont {Zhang}}, \bibinfo {author} {\bibfnamefont {H.-Y.}\ \bibnamefont {Xie}}, \bibinfo {author} {\bibfnamefont {F.}~\bibnamefont {Meng}}, \bibinfo {author} {\bibfnamefont {Y.}~\bibnamefont {Liu}}, \bibinfo {author} {\bibfnamefont {Z.}~\bibnamefont {Bai}}, \bibinfo {author} {\bibfnamefont {S.}~\bibnamefont {Ji}}, \bibinfo {author} {\bibfnamefont {J.}~\bibnamefont {Zhang}}, \bibinfo {author} {\bibfnamefont {K.}~\bibnamefont {Jiang}}, \bibinfo {author} {\bibfnamefont {R.}~\bibnamefont {Zhong}}, \bibinfo {author} {\bibfnamefont {J.}~\bibnamefont {Schneeloch}}, \bibinfo {author} {\bibfnamefont {G.}~\bibnamefont {Gu}}, \bibinfo {author} {\bibfnamefont {L.}~\bibnamefont {Gu}}, \bibinfo {author} {\bibfnamefont {X.}~\bibnamefont {Ma}}, \bibinfo {author} {\bibfnamefont {D.}~\bibnamefont {Zhang}},\ and\ \bibinfo {author} {\bibfnamefont
  {Q.-K.}\ \bibnamefont {Xue}},\ }\bibfield  {title} {\bibinfo {title} {{Presence of $s$-Wave Pairing in Josephson Junctions Made of Twisted Ultrathin ${\mathrm{Bi}}_{2}{\mathrm{Sr}}_{2}{\mathrm{CaCu}}_{2}{\mathrm{O}}_{8+x}$ Flakes}},\ }\href {https://doi.org/10.1103/PhysRevX.11.031011} {\bibfield  {journal} {\bibinfo  {journal} {Phys. Rev. X}\ }\textbf {\bibinfo {volume} {11}},\ \bibinfo {pages} {031011} (\bibinfo {year} {2021})}\BibitemShut {NoStop}%
\bibitem [{\citenamefont {Can}\ \emph {et~al.}(2021)\citenamefont {Can}, \citenamefont {Tummuru}, \citenamefont {Day}, \citenamefont {Elfimov}, \citenamefont {Damascelli},\ and\ \citenamefont {Franz}}]{Can2021}%
  \BibitemOpen
  \bibfield  {author} {\bibinfo {author} {\bibfnamefont {O.}~\bibnamefont {Can}}, \bibinfo {author} {\bibfnamefont {T.}~\bibnamefont {Tummuru}}, \bibinfo {author} {\bibfnamefont {R.~P.}\ \bibnamefont {Day}}, \bibinfo {author} {\bibfnamefont {I.}~\bibnamefont {Elfimov}}, \bibinfo {author} {\bibfnamefont {A.}~\bibnamefont {Damascelli}},\ and\ \bibinfo {author} {\bibfnamefont {M.}~\bibnamefont {Franz}},\ }\bibfield  {title} {\bibinfo {title} {{High-temperature topological superconductivity in twisted double-layer copper oxides}},\ }\href {https://doi.org/10.1038/s41567-020-01142-7} {\bibfield  {journal} {\bibinfo  {journal} {Nature}\ }\textbf {\bibinfo {volume} {17}},\ \bibinfo {pages} {519} (\bibinfo {year} {2021})}\BibitemShut {NoStop}%
\bibitem [{\citenamefont {Ju}\ \emph {et~al.}(2022)\citenamefont {Ju}, \citenamefont {Ren}, \citenamefont {Li}, \citenamefont {Liu}, \citenamefont {Shi}, \citenamefont {Liu}, \citenamefont {Hong}, \citenamefont {Wu}, \citenamefont {Tian}, \citenamefont {Zhou},\ and\ \citenamefont {Xie}}]{Ju2022}%
  \BibitemOpen
  \bibfield  {author} {\bibinfo {author} {\bibfnamefont {L.}~\bibnamefont {Ju}}, \bibinfo {author} {\bibfnamefont {T.}~\bibnamefont {Ren}}, \bibinfo {author} {\bibfnamefont {Z.}~\bibnamefont {Li}}, \bibinfo {author} {\bibfnamefont {Z.}~\bibnamefont {Liu}}, \bibinfo {author} {\bibfnamefont {C.}~\bibnamefont {Shi}}, \bibinfo {author} {\bibfnamefont {Y.}~\bibnamefont {Liu}}, \bibinfo {author} {\bibfnamefont {S.}~\bibnamefont {Hong}}, \bibinfo {author} {\bibfnamefont {J.}~\bibnamefont {Wu}}, \bibinfo {author} {\bibfnamefont {H.}~\bibnamefont {Tian}}, \bibinfo {author} {\bibfnamefont {Y.}~\bibnamefont {Zhou}},\ and\ \bibinfo {author} {\bibfnamefont {Y.}~\bibnamefont {Xie}},\ }\bibfield  {title} {\bibinfo {title} {{Emergence of high-temperature superconductivity at the interface of two Mott insulators}},\ }\href {https://doi.org/10.1103/PhysRevB.105.024516} {\bibfield  {journal} {\bibinfo  {journal} {Phys. Rev. B}\ }\textbf {\bibinfo {volume} {105}},\ \bibinfo {pages} {024516} (\bibinfo {year} {2022})}\BibitemShut
  {NoStop}%
\bibitem [{\citenamefont {Zhao}\ \emph {et~al.}(2023)\citenamefont {Zhao}, \citenamefont {Cui}, \citenamefont {Volkov}, \citenamefont {Yoo}, \citenamefont {Lee}, \citenamefont {Gardener}, \citenamefont {Akey}, \citenamefont {Engelke}, \citenamefont {Ronen}, \citenamefont {Zhong}, \citenamefont {Gu}, \citenamefont {Plugge}, \citenamefont {Tummuru}, \citenamefont {Kim}, \citenamefont {Franz}, \citenamefont {Pixley}, \citenamefont {Poccia},\ and\ \citenamefont {Kim}}]{Zhao2023}%
  \BibitemOpen
  \bibfield  {author} {\bibinfo {author} {\bibfnamefont {S.~Y.~F.}\ \bibnamefont {Zhao}}, \bibinfo {author} {\bibfnamefont {X.}~\bibnamefont {Cui}}, \bibinfo {author} {\bibfnamefont {P.~A.}\ \bibnamefont {Volkov}}, \bibinfo {author} {\bibfnamefont {H.}~\bibnamefont {Yoo}}, \bibinfo {author} {\bibfnamefont {S.}~\bibnamefont {Lee}}, \bibinfo {author} {\bibfnamefont {J.~A.}\ \bibnamefont {Gardener}}, \bibinfo {author} {\bibfnamefont {A.~J.}\ \bibnamefont {Akey}}, \bibinfo {author} {\bibfnamefont {R.}~\bibnamefont {Engelke}}, \bibinfo {author} {\bibfnamefont {Y.}~\bibnamefont {Ronen}}, \bibinfo {author} {\bibfnamefont {R.}~\bibnamefont {Zhong}}, \bibinfo {author} {\bibfnamefont {G.}~\bibnamefont {Gu}}, \bibinfo {author} {\bibfnamefont {S.}~\bibnamefont {Plugge}}, \bibinfo {author} {\bibfnamefont {T.}~\bibnamefont {Tummuru}}, \bibinfo {author} {\bibfnamefont {M.}~\bibnamefont {Kim}}, \bibinfo {author} {\bibfnamefont {M.}~\bibnamefont {Franz}}, \bibinfo {author} {\bibfnamefont {J.~H.}\ \bibnamefont {Pixley}},
  \bibinfo {author} {\bibfnamefont {N.}~\bibnamefont {Poccia}},\ and\ \bibinfo {author} {\bibfnamefont {P.}~\bibnamefont {Kim}},\ }\bibfield  {title} {\bibinfo {title} {{Time-reversal symmetry breaking superconductivity between twisted cuprate superconductors}},\ }\href {https://doi.org/10.1126/science.abl8371} {\bibfield  {journal} {\bibinfo  {journal} {Science}\ }\textbf {\bibinfo {volume} {382}},\ \bibinfo {pages} {1422} (\bibinfo {year} {2023})}\BibitemShut {NoStop}%
\bibitem [{\citenamefont {Li}\ \emph {et~al.}(2019)\citenamefont {Li}, \citenamefont {Lee}, \citenamefont {Wang}, \citenamefont {Osada}, \citenamefont {Crossley}, \citenamefont {Lee}, \citenamefont {Cui}, \citenamefont {Hikita},\ and\ \citenamefont {Hwang}}]{Li2019}%
  \BibitemOpen
  \bibfield  {author} {\bibinfo {author} {\bibfnamefont {D.}~\bibnamefont {Li}}, \bibinfo {author} {\bibfnamefont {K.}~\bibnamefont {Lee}}, \bibinfo {author} {\bibfnamefont {B.~Y.}\ \bibnamefont {Wang}}, \bibinfo {author} {\bibfnamefont {M.}~\bibnamefont {Osada}}, \bibinfo {author} {\bibfnamefont {S.}~\bibnamefont {Crossley}}, \bibinfo {author} {\bibfnamefont {H.~R.}\ \bibnamefont {Lee}}, \bibinfo {author} {\bibfnamefont {Y.}~\bibnamefont {Cui}}, \bibinfo {author} {\bibfnamefont {Y.}~\bibnamefont {Hikita}},\ and\ \bibinfo {author} {\bibfnamefont {H.~Y.}\ \bibnamefont {Hwang}},\ }\bibfield  {title} {\bibinfo {title} {{Superconductivity in an infinite-layer nickelate}},\ }\href {https://doi.org/10.1038/s41586-019-1496-5} {\bibfield  {journal} {\bibinfo  {journal} {Nature}\ }\textbf {\bibinfo {volume} {572}},\ \bibinfo {pages} {624} (\bibinfo {year} {2019})}\BibitemShut {NoStop}%
\bibitem [{\citenamefont {Lu}\ \emph {et~al.}(2021)\citenamefont {Lu}, \citenamefont {Rossi}, \citenamefont {Nag}, \citenamefont {Osada}, \citenamefont {Li}, \citenamefont {Lee}, \citenamefont {Wang}, \citenamefont {Garcia-Fernandez}, \citenamefont {Agrestini}, \citenamefont {Shen}, \citenamefont {Been}, \citenamefont {Moritz}, \citenamefont {Devereaux}, \citenamefont {Zaanen}, \citenamefont {Hwang}, \citenamefont {Zhou},\ and\ \citenamefont {Lee}}]{Lu2021}%
  \BibitemOpen
  \bibfield  {author} {\bibinfo {author} {\bibfnamefont {H.}~\bibnamefont {Lu}}, \bibinfo {author} {\bibfnamefont {M.}~\bibnamefont {Rossi}}, \bibinfo {author} {\bibfnamefont {A.}~\bibnamefont {Nag}}, \bibinfo {author} {\bibfnamefont {M.}~\bibnamefont {Osada}}, \bibinfo {author} {\bibfnamefont {D.~F.}\ \bibnamefont {Li}}, \bibinfo {author} {\bibfnamefont {K.}~\bibnamefont {Lee}}, \bibinfo {author} {\bibfnamefont {B.~Y.}\ \bibnamefont {Wang}}, \bibinfo {author} {\bibfnamefont {M.}~\bibnamefont {Garcia-Fernandez}}, \bibinfo {author} {\bibfnamefont {S.}~\bibnamefont {Agrestini}}, \bibinfo {author} {\bibfnamefont {Z.~X.}\ \bibnamefont {Shen}}, \bibinfo {author} {\bibfnamefont {E.~M.}\ \bibnamefont {Been}}, \bibinfo {author} {\bibfnamefont {B.}~\bibnamefont {Moritz}}, \bibinfo {author} {\bibfnamefont {T.~P.}\ \bibnamefont {Devereaux}}, \bibinfo {author} {\bibfnamefont {J.}~\bibnamefont {Zaanen}}, \bibinfo {author} {\bibfnamefont {H.~Y.}\ \bibnamefont {Hwang}}, \bibinfo {author} {\bibfnamefont {K.-J.}\ \bibnamefont
  {Zhou}},\ and\ \bibinfo {author} {\bibfnamefont {W.~S.}\ \bibnamefont {Lee}},\ }\bibfield  {title} {\bibinfo {title} {{Magnetic excitations in infinite-layer nickelates}},\ }\href {https://doi.org/10.1126/science.abd7726} {\bibfield  {journal} {\bibinfo  {journal} {Science}\ }\textbf {\bibinfo {volume} {373}},\ \bibinfo {pages} {213} (\bibinfo {year} {2021})}\BibitemShut {NoStop}%
\bibitem [{\citenamefont {Hepting}\ \emph {et~al.}(2021)\citenamefont {Hepting}, \citenamefont {Dean},\ and\ \citenamefont {Lee}}]{Hepting2021}%
  \BibitemOpen
  \bibfield  {author} {\bibinfo {author} {\bibfnamefont {M.}~\bibnamefont {Hepting}}, \bibinfo {author} {\bibfnamefont {M.~P.~M.}\ \bibnamefont {Dean}},\ and\ \bibinfo {author} {\bibfnamefont {W.-S.}\ \bibnamefont {Lee}},\ }\bibfield  {title} {\bibinfo {title} {{Soft X-Ray Spectroscopy of Low-Valence Nickelates}},\ }\bibfield  {journal} {\bibinfo  {journal} {Frontiers in Physics}\ }\textbf {\bibinfo {volume} {9}},\ \href {https://doi.org/10.3389/fphy.2021.808683} {10.3389/fphy.2021.808683} (\bibinfo {year} {2021})\BibitemShut {NoStop}%
\bibitem [{\citenamefont {Fan}\ \emph {et~al.}(2024)\citenamefont {Fan}, \citenamefont {LaBollita}, \citenamefont {Gao}, \citenamefont {Khan}, \citenamefont {Gu}, \citenamefont {Kim}, \citenamefont {Li}, \citenamefont {Bhartiya}, \citenamefont {Li}, \citenamefont {Sun}, \citenamefont {Yang}, \citenamefont {Yan}, \citenamefont {Barbour}, \citenamefont {Zhou}, \citenamefont {Cano}, \citenamefont {Bernardini}, \citenamefont {Nie}, \citenamefont {Zhu}, \citenamefont {Bisogni}, \citenamefont {Mazzoli}, \citenamefont {Botana},\ and\ \citenamefont {Pelliciari}}]{Fan2024}%
  \BibitemOpen
  \bibfield  {author} {\bibinfo {author} {\bibfnamefont {S.}~\bibnamefont {Fan}}, \bibinfo {author} {\bibfnamefont {H.}~\bibnamefont {LaBollita}}, \bibinfo {author} {\bibfnamefont {Q.}~\bibnamefont {Gao}}, \bibinfo {author} {\bibfnamefont {N.}~\bibnamefont {Khan}}, \bibinfo {author} {\bibfnamefont {Y.}~\bibnamefont {Gu}}, \bibinfo {author} {\bibfnamefont {T.}~\bibnamefont {Kim}}, \bibinfo {author} {\bibfnamefont {J.}~\bibnamefont {Li}}, \bibinfo {author} {\bibfnamefont {V.}~\bibnamefont {Bhartiya}}, \bibinfo {author} {\bibfnamefont {Y.}~\bibnamefont {Li}}, \bibinfo {author} {\bibfnamefont {W.}~\bibnamefont {Sun}}, \bibinfo {author} {\bibfnamefont {J.}~\bibnamefont {Yang}}, \bibinfo {author} {\bibfnamefont {S.}~\bibnamefont {Yan}}, \bibinfo {author} {\bibfnamefont {A.}~\bibnamefont {Barbour}}, \bibinfo {author} {\bibfnamefont {X.}~\bibnamefont {Zhou}}, \bibinfo {author} {\bibfnamefont {A.}~\bibnamefont {Cano}}, \bibinfo {author} {\bibfnamefont {F.}~\bibnamefont {Bernardini}}, \bibinfo {author} {\bibfnamefont
  {Y.}~\bibnamefont {Nie}}, \bibinfo {author} {\bibfnamefont {Z.}~\bibnamefont {Zhu}}, \bibinfo {author} {\bibfnamefont {V.}~\bibnamefont {Bisogni}}, \bibinfo {author} {\bibfnamefont {C.}~\bibnamefont {Mazzoli}}, \bibinfo {author} {\bibfnamefont {A.~S.}\ \bibnamefont {Botana}},\ and\ \bibinfo {author} {\bibfnamefont {J.}~\bibnamefont {Pelliciari}},\ }\bibfield  {title} {\bibinfo {title} {{Capping Effects on Spin and Charge Excitations in Parent and Superconducting ${\mathrm{Nd}}_{1\ensuremath{-}x}{\mathrm{Sr}}_{x}{\mathrm{NiO}}_{2}$}},\ }\href {https://doi.org/10.1103/PhysRevLett.133.206501} {\bibfield  {journal} {\bibinfo  {journal} {Phys. Rev. Lett.}\ }\textbf {\bibinfo {volume} {133}},\ \bibinfo {pages} {206501} (\bibinfo {year} {2024})}\BibitemShut {NoStop}%
\end{thebibliography}

%

\end{document}


    \title{\com{Supplemental Material for: ``Anisotropic electron damping and energy gap in Bi$_2$Sr$_2$CaCu$_2$O$_{8+\delta}$''}}

    \author{Jiemin Li}
    \affiliation{National Synchrotron Light Source II, Brookhaven National Laboratory, Upton, NY 11973, USA}
    \author{Yanhong Gu}
    \affiliation{National Synchrotron Light Source II, Brookhaven National Laboratory, Upton, NY 11973, USA}
    \author{Takemi Yamada}
    \affiliation{Liberal Arts and Sciences, Toyama Prefectural University, Imizu, Toyama 939-0398, Japan}
    \author{Zebin Wu}
    \affiliation{Condensed Matter Physics and Materials Science Department, Brookhaven National Laboratory, Upton, NY, 11973, USA}
    \author{Genda Gu}
    \affiliation{Condensed Matter Physics and Materials Science Department, Brookhaven National Laboratory, Upton, NY, 11973, USA}
    \author{Tonica Valla}
    \affiliation{Condensed Matter Physics and Materials Science Department, Brookhaven National Laboratory, Upton, NY, 11973, USA}
    \affiliation{\textcolor{black}{Donostia International Physics Center, E-20018 Donostia-San Sebastian, Spain}}
    \author{Ilya Drozdov}
    \affiliation{Condensed Matter Physics and Materials Science Department, Brookhaven National Laboratory, Upton, NY, 11973, USA}
    \author{Ivan Bo\v{z}ovi\'{c}}
    \affiliation{Condensed Matter Physics and Materials Science Department, Brookhaven National Laboratory, Upton, NY, 11973, USA}
    \author{Mark P. M. Dean}
    \affiliation{Condensed Matter Physics and Materials Science Department, Brookhaven National Laboratory, Upton, NY, 11973, USA}
    \author{Takami Tohyama}
    \affiliation{Department of Applied Physics, Tokyo University of Science, Katsushika, Tokyo 125-8585, Japan}
    \author{Jonathan Pelliciari}
    \affiliation{National Synchrotron Light Source II, Brookhaven National Laboratory, Upton, NY 11973, USA}
    \author{Valentina Bisogni}
    \affiliation{National Synchrotron Light Source II, Brookhaven National Laboratory, Upton, NY 11973, USA}

    \maketitle
    
    This supplemental material contains: I. experimental details; II. calculation of charge susceptibility; III. determination of scattering rate; IV. RIXS simulation; V. \textit{\textbf{Q}}-dependence of elastic intensity change.
    
    \section{Experimental Details.}
     \textit{Experimental details.}---High-quality single crystals of nearly optimally doped Bi$_2$Sr$_2$CaCu$_2$O$_{8+\delta}$ ($T_c$=91K, \com{$T^*$$\sim$190K})~\cite{Hashimoto2014} were grown by the traveling-solvent floating-zone method~\cite{Wen2008}. The samples were characterized by ARPES as described in previously published works~\cite{Drozdov2018, Valla2020}. A fresh surface was prepared in air by cleaving with scoth-tape, just before loading the sample in the vacuum chamber. The RIXS measurements were performed at the SIX 2-ID beamline of the National Synchrotron Light Source II (NSLS-II)~\cite{Dvorak16} with an energy resolution of $\Delta E$ $\sim$ 35~meV (full width at half maximum) at the Cu $L_3$ edge. \textcolor{black}{The momentum resolution was estimated to be $\sim$~0.007~r.l.u. by considering the angular acceptance of the RIXS spectrometer.} The incident photon energy was tuned to the maximum of Cu $L_3$ absorption peak.  Linear-vertical polarization was used to maximize the charge contribution in the RIXS spectra~\cite{Minola2015}. The sample was mounted with the surface normal [001] and the [110] axis lying in the scattering plane. Throughout the experiment, the scattering angle was fixed to 150$^{\circ}$. The Miller indices in this study are defined by a pseudotetragonal unit cell, with $a=b=3.82$\AA and $c=30.7$\AA~\cite{Chaix2017}. The momentum transfer \textit{\textbf{Q}} is defined in reciprocal lattice units (r.l.u.) as \textit{\textbf{Q}} = $H$\textit{\textbf{a}}$^*$ + $K$\textit{\textbf{b}}$^*$ + $L$\textit{\textbf{c}}$^*$ where $|$\textit{\textbf{a}}$^*$$|$=$2\pi/a$, etc.

    \section{Charge susceptibility calculation}
        It has been shown that \textcolor{black}{the RIXS cross-section is proportional to the charge dynamic structure factor, $I^\mathrm{RIXS}\propto S_c(\textit{\textbf{Q}}, \omega)$ = Im$\chi_c(\textit{\textbf{Q}}, \omega)/[1-e^{-\omega/(k_BT)}]$~\cite{Ament2011, Marra2013, Jia2016}, where \textit{\textbf{Q}} and $\omega$ are respectively the momentum and energy transfer from the photon to the material. Here, we firstly calculate the imaginary of charge susceptibility Im$\chi_c(\textit{\textbf{Q}}, \omega)$} as a function of both momentum and energy. The $\chi_c(\textit{\textbf{Q}}, \omega)$ for superconducting (SC) and normal (N) states is expressed respectively as:
       \begin{widetext}
            \begin{equation}
                \begin{aligned}
                    \chi_{c}^\mathrm{SC}(\bi{Q},\omega) =& \frac{1}{2N} \sum_\bi{k} \Biggl\{\frac{1+\Omega_\bi{k,Q}}{2} \Biggr[\frac{f_{\bi{k}+\bi{Q}}-f_\bi{k}}{\omega+i \Gamma^\mathrm{SC}_{\bi{Q},\bi{k}}(\omega)+E_\bi{k}-E_{\bi{k}+\bi{Q}}} + \frac{f_\bi{k}-f_{\bi{k}+\bi{Q}}}{\omega+i \Gamma^\mathrm{SC}_{\bi{Q},\bi{k}}(\omega)+E_{\bi{k}+\bi{Q}}-E_\bi{k}}\Biggr]\\ &+  \frac{1-\Omega_\bi{k,Q}}{2} \Biggr[ \frac{1-f_{\bi{k}+\bi{Q}}-f_\bi{k}}{\omega+i\Gamma^\mathrm{SC}_{\bi{Q},\bi{k}}(\omega)+E_\bi{k}+E_{\bi{k}+\bi{Q}}} + \frac{f_{\bi{k}+\bi{Q}}+f_\bi{k}-1}{\omega+i\Gamma^\mathrm{SC}_{\bi{Q},\bi{k}}(\omega)-E_\bi{k}-E_{\bi{k}+\bi{Q}}}
                    \Biggr]
                    \Biggl\},
                \end{aligned}
                \label{Eq1}
            \end{equation}
        \end{widetext}

        \begin{widetext}
            \begin{equation}
                \begin{aligned}
                    \chi_{c}^\mathrm{N}(\bi{Q},\omega) = &\frac{1}{2N} \sum_\bi{k} \Biggl\{ \frac{1+\eta_\bi{k,Q}}{2} \Biggr[ \frac{f_{\bi{k}+\bi{Q}}-f_\bi{k}}{\omega+i\Gamma^\mathrm{N}(\omega)+|\varepsilon_\bi{k}|-|\varepsilon_{\bi{k}+\bi{Q}}|} + \frac{f_\bi{k}-f_{\bi{k}+\bi{Q}}}{\omega+i\Gamma^\mathrm{N}(\omega)+|\varepsilon_{\bi{k}+\bi{Q}}|-|\varepsilon_\bi{k}|}  \Biggr] \\ &+  \frac{1-\eta_\bi{k,Q}}{2} \Biggr[ \frac{1-f_{\bi{k}+\bi{Q}}-f_\bi{k}}{\omega+i\Gamma^\mathrm{N}(\omega)+|\varepsilon_\bi{k}|+|\varepsilon_{\bi{k}+\bi{Q}}|} + \frac{f_{\bi{k}+\bi{Q}}+f_\bi{k}-1}{\omega+i\Gamma^\mathrm{N}(\omega)-|\varepsilon_\bi{k}|-|\varepsilon_{\bi{k}+\bi{Q}}|}
                    \Biggr]
                    \Biggl\},
                \end{aligned}
                \label{Eq2}
            \end{equation}
        \end{widetext}
        where $\Omega_\bi{k,Q} = (\varepsilon_\bi{k}\varepsilon_{\bi{k}+\bi{Q}}-\Delta_\bi{k}\Delta_{\bi{k}+\bi{Q}})/{E_\bi{k}E_{\bi{k}+\bi{Q}}}$ with $E_\bi{k}=\sqrt{\varepsilon_\bi{k}^2+\Delta_\bi{k}^2}$ and the $d$-wave gap $\Delta_\bi{k} = \frac{\com{\Delta}}{2}[\cos(k_x)-\cos(k_y)]$, $\eta_\bi{k,Q}=\varepsilon_\bi{k}\varepsilon_{\bi{k}+\bi{Q}}/|\varepsilon_\bi{k}\varepsilon_{\bi{k}+\bi{Q}}|$. The $\varepsilon_\bi{k}$ is the tight binding band structure of the normal state \ta{taken from Ref.~\cite{Norman1995}} and \ta{the $f_\bi{k}=f(\epsilon_\bi{k})$ is the Fermi distribution function with $\epsilon_\bi{k}=E_\bi{k}$ for Eq.~(\ref{Eq1}) and $\epsilon_\bi{k}=|\varepsilon_\bi{k}|$ for Eq.~(\ref{Eq2}).} The scattering rate $\Gamma$ characterizes the microscopic electron interactions~\cite{Sobota2021}. In the N state, it is approximated as only a function of electron energy~\cite{Schlesinger1990, Bogdanov2002}, i.e., \ta{a marginal Fermi liquid form} $\Gamma^\mathrm{N}(\omega)=\delta + \omega$ with \ta{a constant term} $\delta=0.002$~eV. In the SC state, instead, multiple interactions (such as electron-phonon and electron-electron interactions) prevail in the system, therefore introducing a non-trivial energy and momentum dependent $\Gamma$~\cite{Sobota2021}. As shown in Eq.~(1) of main text, our $\Gamma^\mathrm{SC}_{\bi{Q},\bi{k}}(\omega)$ implements the anisotropic $\textit{\textbf{k}}$-dependence that had been previously reported in cuprates by ARPES~\cite{Valla1999,Valla2000,Abdel2006, Kaminski2005, Chang2013}, essential for evaluating the charge susceptibility in \textit{\textbf{Q}}-space $\chi_c(\textit{\textbf{Q}}, \omega)$.

        \ta{To perform the calculations, we considered $1024\times 1024$ \bi{k} points in the first Brillouin zone of a square lattice and divided the energy interval of $[0,0.5]$~eV into 1000 meshes. The temperatures for the SC and N states are assumed to be 1.2~K and 120~K, respectively.}

        \begin{figure}[t]
            \includegraphics[width=0.9\columnwidth]{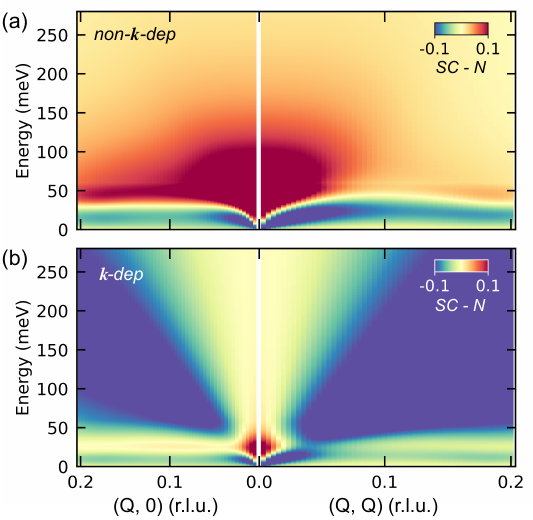}
            \caption{Spectral difference of Im$\chi_c(\textit{\textbf{Q}}, \omega)$ between the SC and N states in momentum-energy space. (a, b) are the results of calculations done without and with the \ta{\bi{k}-dependent anisotropic} scattering rate, respectively.
            }
            \label{Fig_sm1}
        \end{figure}

        Figure~1(c, d, e) in main text summarizes the calculated Im$\chi_{c}^\mathrm{SC}(\textit{\textbf{Q}}, \omega)$ without/with the anisotropic $\textit{\textbf{k}}$-dependence in $\Gamma^\mathrm{SC}_{\bi{Q},\bi{k}}(\omega)$ and the Im$\chi_{c}^\mathrm{N}(\textit{\textbf{Q}}, \omega)$. As discussed in the main text, Im$\chi_{c}^\mathrm{N}(\textit{\textbf{Q}}, \omega)$ evolves from 0 meV at the Brillouin Zone (BZ) center to a few hundreds meV at higher \textit{\textbf{Q}}-positions in the normal state. In the SC state the opening of \com{the electronic energy} gap pushes Im$\chi_{c}^\mathrm{SC}(\textit{\textbf{Q}}, \omega)$ below \com{$\Delta$} ($\sim 30$ meV) to higher energy, see Figs.~1 (c, d) in the main text. As a consequence, the spectral differences of Im$\chi_c(\textit{\textbf{Q}}, \omega)$ between the SC and N states display noticeable changes in the low-\textit{\textbf{Q}} region [$|\textit{\textbf{Q}}| < 0.1$ (r.l.u.)], as demonstrated in Fig.~\ref{Fig_sm1}. The difference between (a) and (b) in Fig.~\ref{Fig_sm1} is the inclusion of $\textit{\textbf{k}}$-dependent anisotropic scattering rate (\textit{\textbf{k}}-DAES) in the calculation of Im$\chi_{c}^\mathrm{SC}(\textit{\textbf{Q}}, \omega)$. Without \textit{\textbf{k}}-DAES, the spectral difference presents a dip-peak feature in the displayed $\textit{\textbf{Q}}$-region [see Fig.~\ref{Fig_sm1}(a)], while it mostly displays a dip-only feature when accounting for \textit{\textbf{k}}-DAES [see Fig.~\ref{Fig_sm1}(b)]. Besides, we notice that the contrast between these two scenarios is more pronounced along the nodal direction-(Q, Q) than the anti-nodal one. Based on this guideline, the RIXS measurements were performed along the nodal direction. \jl{Note that for very small \textit{\textbf{Q}} values ($|$\textit{\textbf{Q}}$|$ smaller than 0.01), we observe a dip-peak-like structure in {\it{SC-N}} for both scenarios, due to the highly coherent nature of the charge excitations close to the BZ center}. \vb{For this reason, we comment that our finding -- importance of including \textit{\textbf{k}}-DAES -- is compatible with the Raman results presented in Ref.~\cite{Suzuki2018}, where the authors observed a dip-peak like shape in their $SC$-$N$ spectrum at  \textit{\textbf{Q}} $\sim 0$ (as it is the case for Raman)}.

    \section{RIXS simulation}
        
        \begin{figure}[t]
            \includegraphics[width=0.7\columnwidth]{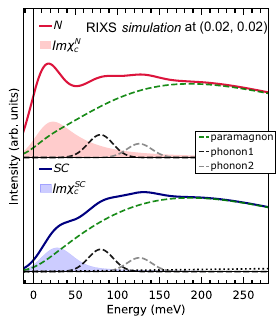}
            \caption{RIXS simulations for the SC and N states. The blue and red lines are the simulated spectra in the SC and normal states, respectively. The color-filled areas are the corresponding Im$\chi_c(\textit{\textbf{Q}}, \omega)$.The green dashed line is paragmanon and the black/gray dashed lines indicate two phonon modes.
            }
            \label{Fig_sm2}
        \end{figure}

        \begin{figure*}
            \includegraphics[width=0.85\textwidth]{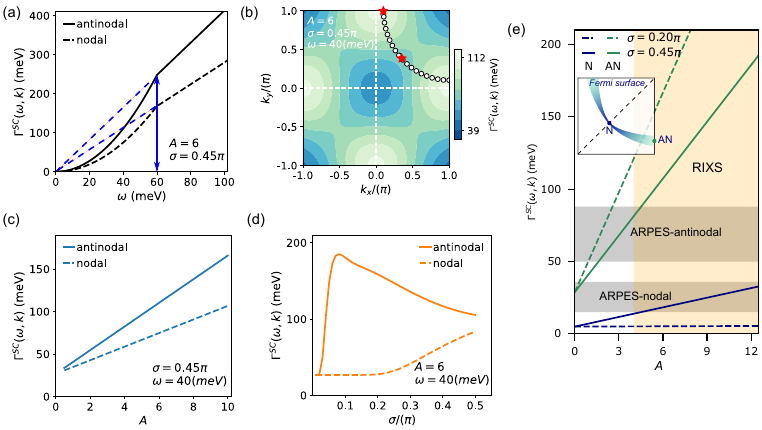}
            \caption{Scattering rate $\Gamma^\mathrm{SC}_{\bi{Q},\bi{k}}(\omega)$ at \textit{\textbf{Q}}=0 in SC state. (a) Energy dependent behavior of the scattering rate evaluated at the nodal and anti-nodal points [see red stars in (b)], with $A=6$ and $\sigma=0.45\pi$. The line with double arrows indicates the coherent energy gap size of \ta{\com{2$\Delta=60$}}~meV. The blue dashed line is the linear interpolation (down to 0~meV) of scattering rate below $\omega$=60~meV. (b) Scattering rate vs momentum $\textit{\textbf{k}}$ at an energy cut of $\omega$=40~meV, with $A=6$ and $\sigma=0.45\pi$. The white dots denote the Fermi arc and the stars indicate the two $\textit{\textbf{k}}$-positions (nodal and anti-nodal points) examined in the (a, c, d, e) panels. (c) Scattering rate variation as a function of $A$, at $\omega=40$~meV and $\sigma=0.45\pi$. (d) Scattering rate variation as a function of $\sigma$, at $\omega=40$~meV and $A=6$. (e) Scattering rate as a function of $A$ with $\sigma=0.2\pi$ (dashed lines) and $\sigma=0.45\pi$ (solid lines). The gray areas indicate the region of scattering rate deduced from ARPES data~\cite{Valla1999, Valla2000}. The two lines are calculated from Eq.~(1) in main text with $\sigma=0.45\pi$, at two representing positions (N: nodal point; AN: anti-nodal point). The yellow area is the region supported by the RIXS data. The inset depicts a quadrant of the Brillioun zone with the arc denoting the Fermi surface of cuprates.
            }
            \label{Fig_sm3}
        \end{figure*}

        We simulate the low-energy portion of the RIXS spectra as the sum of the charge, phonons, and paramagnons~\cite{Dean2013, Chaix2017}, i.e., 
        \begin{equation*}
            I_\mathrm{RIXS} = \textcolor{black}{S_c(\textit{\textbf{Q}}, \omega)}+ C_1 \times I_\mathrm{ph1}+ C_2 \times I_\mathrm{ph2}+ C_3 \times I_\mathrm{paramagnon}
            \label{Eq3}
        \end{equation*}
        with the intensities of phonons and paramagnons adjusted by the parameters $C_1, C_2, C_3$. The charge component is modelled with the charge dynamic structure factor  $\textcolor{black}{S_c(\textit{\textbf{Q}}, \omega)}$. As discussed in the main text, the peak at $\sim$80 meV is attributed to the apical phonon mode (phonon1) and the one at $\sim$125 meV to a combined phonon mode (phonon2) between the apical and the A$_{1g}$ phonons~\cite{Devereaux2016}. These two phonons barely change in energy in our measured \textit{\textbf{Q}}-positions. Beside, it had been also shown in~\cite{Devereaux2016} that only these two phonon modes contribute to the RIXS intensity when approaching the BZ center. In the simulation, we represent these two phonon modes with Gaussian curves and fix their energies to 80 meV and 125 meV, respectively. Their widths are also constrained to instrument energy resultion, i.e., $\Delta E\sim30$ meV, see black and gray dashed lines in Fig.~\ref{Fig_sm2}.  The paramagnon is heavily damped in doped cuprates~\cite{Robarts2019} and strongly overlaps with phonons in the low-\textit{\textbf{Q}} region. As reported in~\cite{Dean2013,Robarts2019}, the energy of the paramagnon can be approximated as a linear function of \textit{\textbf{Q}} when approaching the BZ center. We thus use an anti-Lorentzian shape~\cite{Jiemin2020} (green dashed line in Fig.~\ref{Fig_sm2}) to denote the paramagnon with the energy approximately following a relation of $E=|\textit{\textbf{Q}}| \times 1000$ (meV). The width of paramagnon was fixed to 320 meV, comparable to reported ones~\cite{Dean2013,Chaix2017}. To account for the temperature effect, \textcolor{black}{all these phonons and paramagnons had been corrected by the Bose factor~\cite{Robarts2019}.} 
        In addition, the intensities of simulated phonons and paramagnons had also been scaled through the parameters $C_1, C_2, C_3$ to match the calculated $S_c(\textit{\textbf{Q}}, \omega)$. The final RIXS simulations were then convoluted with the instrument energy resolution before making the comparison with RIXS experimental data.

    \section{Determination of the scattering rate in SC state}
    The scattering rate $\Gamma^\mathrm{SC}_{\bi{Q},\bi{k}}(\omega)$ [see Eq.~(1) in main text] is expressed as the product of two terms: the first one is the energy distribution with a cutoff energy set by \com{$2\Delta$}, and the second one is a phenomenological description of its anisotropic momentum dependence. To check how $\Gamma^\mathrm{SC}_{\bi{Q},\bi{k}}(\omega)$ varies with different parameters, we summarized the calculations of $\Gamma^\mathrm{SC}_{\bi{Q},\bi{k}}(\omega)$ vs. $\omega$, $k$, $A$ and $\sigma$ at \textit{\textbf{Q}}=0 in Fig.~\ref{Fig_sm3}. In Fig.~\ref{Fig_sm3}(a), we display the energy behavior of $\Gamma^\mathrm{SC}_{\bi{Q},\bi{k}}(\omega)$ at two $\textit{\textbf{k}}$ positions, the nodal and the antinodal points. At the nodal point, the $\Gamma^\mathrm{SC}_{\bi{Q},\bi{k}}(\omega)$ increases linearly with electron energy $\omega$ (black dashed line). Instead, at the antinodal point, it varies linearly with $\omega$ for energies higher than 60~meV (that is \com{2$\Delta$}) while it turns into a parabolic shape for smaller energies (solid black line). This behaviour is consistent with  ARPES data from cuprates~\cite{Chang2013}. In Fig.~\ref{Fig_sm3}(b), we examine the evolution of $\Gamma^\mathrm{SC}_{\bi{Q},\bi{k}}(\omega)$ in $\textit{\textbf{k}}$-space, at a fixed energy $\omega$=40~meV (above $\Delta_\mathrm{SC}$). As introduced in Fig.~1(b) in main text, we use eight Gaussian curves centered at antinodal points to account for the \textit{\textbf{k}}-DAES in $\Gamma^\mathrm{SC}_{\bi{Q},\bi{k}}(\omega)$. The resulting $\Gamma^\mathrm{SC}_{\bi{Q},\bi{k}}(\omega)$ presents minima at nodal points while it maximizes at the anti-nodal points and zone center [see Fig.~\ref{Fig_sm3}(b)], compatibly with the momentum dependence studied by ARPES~\cite{Valla2000, Chang2013}. Therefore, our proposed scattering rate captures the electron behavior in both energy and momentum space~\cite{Ivan1991,Cooper1992,Bonn1992, Rieck1995,Valla2000, Chang2013}.

    In the phenomenological form of $\Gamma^\mathrm{SC}_{\bi{Q},\bi{k}}(\omega)$ including the \textit{\textbf{k}}-DAES, there are two free parameters, $A$ and $\sigma$, which respectively characterize the amplitude and width of the Gaussian curves introduced to describe the momentum dependence of $\Gamma^\mathrm{SC}_{\bi{Q},\bi{k}}(\omega)$, see Fig.~1(b). In Figs.~\ref{Fig_sm3}(c, d), we check how the $\Gamma^\mathrm{SC}_{\bi{Q},\bi{k}}(\omega)$ evolves when changing the $A$ and $\sigma$. At a fixed $\sigma$ value, e.g., $\sigma=0.45\pi$, the $\Gamma^\mathrm{SC}_{\bi{Q},\bi{k}}(\omega)$ shows a linear dependence on $A$ (at both the nodal and antinodal points), with the absolute value of the scattering rate being larger at the antinodal point, see Fig.~\ref{Fig_sm3}(c). At a given $A$, i.e., $A = 6$, $\Gamma^\mathrm{SC}_{\bi{Q},\bi{k}}(\omega)$ behaves instead differently at different positions in $\textit{\textbf{k}}$-space. At the antinodal point, $\Gamma^\mathrm{SC}_{\bi{Q},\bi{k}}(\omega)$ quickly increases and then gradually decreases when enlarging $\sigma$. At the nodal point, the value of $\Gamma^\mathrm{SC}_{\bi{Q},\bi{k}}(\omega)$ is overall much smaller than at the antinodal point, remaining almost invariant up to $\sigma\sim$0.25$\pi$, while it increases afterward, see Fig.~\ref{Fig_sm3}(d). 

    To obtain appropriate values for $A$ and $\sigma$, we firstly compare $\Gamma^\mathrm{SC}_{\bi{Q},\bi{k}}(\omega)$ to the values extracted from previous ARPES data on Bi$_2$Sr$_2$CaCu$_2$O$_{8+\delta}$~\cite{Valla1999, Valla2000}, see the gray shaded bands reported in Fig.~\ref{Fig_sm3}(e) for the nodal and antinodal points. When considering a small $\sigma$ value, i.e. $\sigma$=0.2$\pi$, the corresponding $\Gamma^\mathrm{SC}_{\bi{Q},\bi{k}}(\omega)$ evaluated at the nodal point (dark-blue dashed line) is always smaller than the scattering rate reported by the ARPES at different $A$'s; while at the antinodal point (green dashed line), it overlaps with the ARPES data within the $A$ window that goes from  1 to 3, roughly. Since at this $\sigma$ value the calculated $\Gamma^\mathrm{SC}_{\bi{Q},\bi{k}}(\omega)$ does not fully satisfy the experimental observations, we examine a larger $\sigma$. At $\sigma$=0.45$\pi$, the $\Gamma^\mathrm{SC}_{\bi{Q},\bi{k}}(\omega)$ at the nodal point agrees with the ARPES values for $A \geq 4.5$ (dark-blue solid line), as the larger $\sigma$ value causes an increase of scattering rate as discussed in Fig.~\ref{Fig_sm3}(d). Rather, at antinodal point, the larger $\sigma$ value suppresses $\Gamma^\mathrm{SC}_{\bi{Q},\bi{k}}(\omega)$, shifting the $A$ window that satisfies the ARPES data towards larger values, starting from $A$=2. Based on this comparison, we fix the $\sigma$ value to 0.45$\pi$ for our calculations, as it satisfies both the nodal and anti-nodal experimental observations from ARPES. 

\begin{figure}[t]
            \includegraphics[width=0.65\columnwidth]{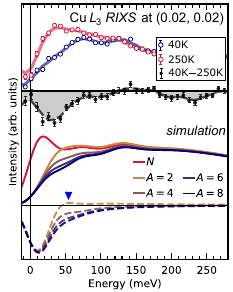}
            \caption{Comparison of RIXS data (top panel) and simulations (bottom panel) done for different $A$ values at a fixed $\sigma=0.45\pi$, for $\textit{\textbf{Q}}$=(0.02, 0.02).
            }
            \label{Fig_sm4}
        \end{figure}

    With such a $\sigma$ value, we refine next the $A$ parameter by referring to our RIXS data. We calculate the Im$\chi_{c}(\textit{\textbf{Q}}, \omega)$ using $\Gamma^\mathrm{SC}_{\bi{Q},\bi{k}}(\omega)$ from different $A$'s, and then simulate the RIXS spectra following the procedure described in Sec.~II. As shown in Fig.~\ref{Fig_sm4}, the hump structure in the SC state keeps reducing when increasing the $A$ value. For a small $A$ value, i.e. $A=2$ (green solid line), the hump structure below 50~meV is stronger than the spectral weight calculated for the normal state (red solid line), thus introducing a dip-peak feature in the spectral difference (green dashed line in Fig.~\ref{Fig_sm4}). When increasing the $A$ value to 4 or higher, the hump structure in the SC spectrum gets suppressed, yielding a simple dip feature in the $SC$-$N$ spectral difference, similar to the RIXS experimental data reproduced in the top panel of Fig.~\ref{Fig_sm4}. Therefore, for $A \geq 4$, we achieve a qualitative good agreement between the calculated and measured RIXS spectral differences [refer to the yellow area reproduced in Fig.~\ref{Fig_sm3}(e)]. Combining both the ARPES and RIXS analysis, our study converges on the following selection of parameters: $\sigma$=0.45$\pi$ and 4$\leq A \leq$6. For an optimal match with the RIXS spectra, we fix $A$=6 in the main text. In conclusion, given the sensitivity of the RIXS spectra to the scattering rate, we corroborate the use of RIXS as a complementary probe to ARPES for the study of the electron dynamics in momentum space.


        \begin{figure}[t]
            \includegraphics[width=1.0\columnwidth]{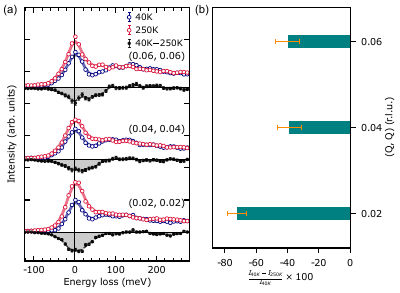}
            \caption{Quasi-elastic peak intensity at different $\textit{\textbf{Q}}$-positions, in SC and normal state, in Bi$_2$Sr$_2$CaCu$_2$O$_{8+\delta}$ ($T_c$ = 91 K). (a) Cu $L_3$ RIXS data at 40K (blue dots) and at 250K (red dots), together with their spectral difference (black dots). The solid lines are  the corresponding smoothed results. (b) Percentage of intensity change of the quasi elastic line in SC and normal state, as a function of momentum. The histogram is obtained by integrating the quasi elastic line within an energy window of [-30meV, +30meV].
            }
            \label{Fig_sm5}
        \end{figure}

        \begin{figure}[t]
            \includegraphics[width=1.0\columnwidth]{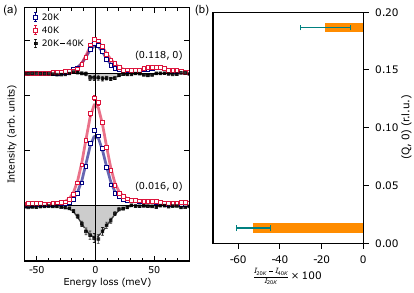}
            \caption{Quasi-elastic peak intensity at different $\textit{\textbf{Q}}$-positions, in SC and normal state, in La$_{2-x}$Sr$_x$CuO$_4$ with $x\sim0.16$ ($T_c = 38$K). (a) O $K$-RIXS data at 20K (blue dots) and at 40K (red dots), together with their spectral difference (black dots). The solid lines are  the corresponding smoothed results. (b) Percentage of intensity change of the quasi elastic line in SC and normal state, as a function of momentum. The histogram is obtained by integrating the quasi elastic line within an energy window of [-20meV, +20meV]. Note that the energy resolution of RIXS instrument at O $K$-edge is $\Delta E\sim17$ meV.
            }
            \label{Fig_sm6}
        \end{figure}

    \section{Elastic peak intensity and closure of the energy gap} To understand the momentum evolution of the low-energy RIXS spectral weight as approaching the BZ center, we examine in this section the changes observed in the elastic peak intensity at several \textit{\textbf{Q}}-positions, both in the $SC$ and normal states. Fig.~\ref{Fig_sm5}(a) reports the raw RIXS spectra measured on Bi$_2$Sr$_2$CaCu$_2$O$_{8+\delta}$, zoomed in the low-energy sector. From this plot, it clearly emerges that the elastic peak intensity measured in the SC state (blue dotted line) is heavily suppressed with respect to the normal state (red dotted line) at \textit{\textbf{Q}}=(0.02, 0.02), while their difference gets quickly reduced at larger \textit{\textbf{Q}}'s. Fig.~\ref{Fig_sm5}(b) summarizes this observation by presenting the \textit{\textbf{Q}} evolution of the elastic spectral difference (expressed in percentage), evaluated by integrating the $SC$-$N$ quantity in the $\pm 30$ meV energy range. When moving away from the BZ center, the elastic peak difference between superconducting and normal states quickly evolves from $\sim -70\%$ at \textit{\textbf{Q}}=(0.02, 0.02) to $\sim -40\%$ at \textit{\textbf{Q}}=(0.04, 0.04) or (0.06, 0.06). Such a sudden change is consistent with the gap closure in the normal state and the strong dispersion of $\chi_c(\textit{\textbf{Q}}, \omega)$ as \textit{\textbf{Q}} is increased, see Figs.~1(c, d, e) in main text. This result further supports the sensitivity of the RIXS spectral weight to the opening and closure of the \com{electronic energy} gap. 
    
    To prove the generality of this statement across the cuprates family, we report similar observations for La$_{2-x}$Sr$_x$CuO$_4$ ($T_c$ = 38~K \cite{Yoshida2012, He2023}), see Fig.~\ref{Fig_sm6}. In this case, the RIXS data displayed in Fig.~\ref{Fig_sm6}(a) were measured at the O $K$-edge, while exciting on the maximum of the first XAS pre-peak \cite{Chen1992}. Due to the intrinsically better energy resolution ($\Delta E\sim17$ meV) achieved at this edge, the histogram in Fig.~\ref{Fig_sm6}(b) is realized by integrating within the energy range of $\pm$20 meV. Similarly to the results of Bi$_2$Sr$_2$CaCu$_2$O$_{8+\delta}$, also in the La$_{2-x}$Sr$_x$CuO$_4$ case we observe a strong change in the elastic spectral difference between SC and N states, as \textit{\textbf{Q}} approaches the BZ center, due to the closure of the \com{electronic energy} gap. \textcolor{black}{Note that owing to multiple phonons appearing in the low-energy region of O $K$-edge RIXS spectra from La$_{2-x}$Sr$_x$CuO$_4$~\cite{Jiemin2020, Li2023}, it is still challenging to achieve a direct comparison with our RIXS simulations. A better energy resolution at O $K$-edge than the 17~meV used for our study would be beneficial for analyzing the electronic effect discussed in our work.}

    
%